\documentclass[12pt]{article}
\usepackage{amssymb}
\usepackage{amsmath}
\usepackage[space]{cite}
\usepackage{graphicx}
\usepackage{tcolorbox}
\usepackage{float}

\usepackage{hyperref}
\numberwithin{equation}{section}

\newcommand{\be}{\begin{equation}}
\newcommand{\ee}{\end{equation}}

\newcommand{\tr}{\mathop{\rm tr}\nolimits}

\newcommand{\bI}{\mathbb{I}}
\newcommand{\cH}{\mathcal{H}}
\newcommand{\cL}{\mathcal{L}}

\newcommand{\cT}{\mathcal{T}}

\newcommand{\cA}{\mathcal{A}}
\newcommand{\cB}{\mathcal{B}}

\newcommand{\cD}{\mathcal{D}}

\begin{document}

\begin{titlepage}
\vspace{.5in}
\begin{center}

{\LARGE Introduction to classical and quantum integrability}\\
\vspace{1in}
\large 

Ana L. Retore  \footnote{School of Mathematics \& Hamilton 
Mathematics Institute, Trinity College Dublin, Dublin, Ireland, retorea@tcd.ie}

retorea@tcd.ie

\end{center}

\vspace{.5in}

\begin{abstract}
	In these lecture notes we aim for a pedagogical introduction to both classical and quantum integrability. Starting from Liouville integrability and passing through Lax pair and r-matrix we discuss the construction of the conserved charges for classical integrable models taking as example the harmonic oscillator. The construction of these charges for 2D integrable field theories is also discussed using a Lax connection and the Sine-Gordon model as example. On the quantum side, the XXZ spin chain is used to explain the systematic construction of the conserved charges starting from a quantum R-matrix, solution of the quantum Yang-Baxter equation. The diagonalization of these charges is performed using the algebraic Bethe ansatz. At the end, the interpretation of the R-matrix as an S-matrix in a scattering process is also presented.
	
	\
	\noindent These notes were written for the lectures delivered at the school ``Integrability, Dualities and Deformations", that ran from 23 to 27 August 2021 in Santiago de Compostela and virtually.
	
\end{abstract}

\end{titlepage}

\setcounter{footnote}{0}

\tableofcontents

\newpage

\section{Introduction and motivation}\label{sec:intro}

Integrable models play a role in many areas of physics ranging from condensed matter, to string theory, passing through Temperley-Lieb and Hecke algebras, quantum groups and Yangians, the bootstrap program, AdS/CFT, sigma models, quantum computing, statistical mechanics and many others\cite{Hubbard_1965RSPSA,Shastry_1986,Turaev:1988eb,Jimbo:1989qm,Wu_1993,Jimbo:1994,Chari:1994pz,essler2005one,abramsky2009temperleylieb,Crampe:2010cs,Beisert:2010jr,Hoare:2011wr,Hoare:2015kla,Borsato:2013qpa,Batchelor:2015osa,Demulder:2017zhz,Foerster2018,Delduc:2018hty,Padmanabhan:2019qed,Alcaraz:2020rru,Sa:2020bdo,VanDyke:2021nuz}. Three aspects of integrable models are explored in the \textit{Integrability, dualities and deformations school}: the basic concepts and techniques on both classical and quantum integrability presented in these lecture notes, integrable deformations of sigma models in Ben Hoare's lecture notes \cite{Hoare:2021dix} and 4-dimensional Chern-Simons theory and integrable field theories in Sylvain Lacroix's lecture notes\cite{Lacroix:2021iit}.  

The reason why integrable models became so widely studied is that they possess a large amount of symmetries, which make them highly constrained and by consequence ``exactly" solvable.

It is very hard to know \textit{a priori} if a model is integrable or not. Several famous models like Kepler's problem, harmonic oscillator, KdV (Korteweg de-Vries), mKdV (modified KdV), Sine(Sinh)-Gordon and non-linear Schrödinger (NLS) are actually integrable \cite{Faddeevbooks,babelon_bernard_talon_2003,Arutyunov2019}.  Many of them are however highly non-linear and, without extra information, it would not be easy to guess that they are integrable. 

They have nonetheless a lot of very interesting hidden structures. Solitons solutions are one example. They are localized solutions that preserve their form while moving, and in a scattering they emerge changed only by a phase shift. There are for example, systems called classical integrable hierarchies which have an infinite number of non-linear integrable equations, and all of them have soliton solutions of the same form.

We are interested however, in discussing what these models have in common and which strategy we can use to find and understand them. 

One of the first very useful tools introduced in this area was the definition of Liouville integrability and the Liouville theorem. With them we learned that if a system with a finite number of degrees of freedom, in a $ 2n $-dimensional phase space, has $ n $ charges in involution, it is always possible to solve such system by performing a finite number of integrals. 

With the introduction of the Lax pair and the classical Yang-Baxter equation, the systematic construction of several new integrable models became possible, including integrable hierarchies associated to some algebras. The Lax connection plays a similar role for 2D integrable field theories. 

The construction of new integrable models remains a modern and challenging task where several recent advancements have been made.
There is a particular class of models that has been widely studied and where a lot of progress has been achieved in this direction: the 2-dimensional  sigma models and its deformations. The Lax connection appears very naturally in these models and in both \cite{Hoare:2021dix} and \cite{Lacroix:2021iit} such cases are discussed using the Principal Chiral model (PCM) and its deformations as examples. In \cite{Lacroix:2021iit} we learn that different integrable sigma models can appear depending on the choice of boundary conditions in a 4-dimensional Chern-Simons theory \cite{Costello:2013zra,Witten:2016spx,Costello:2017dso,Costello:2018gyb}. So, this 4-dimensional Chern-Simons model works as a method to construct integrable models, which given the usual difficulties already discussed in doing this, makes it even more valuable.  

Models that are classically integrable and therefore have many conserved charges, not necessarily remain integrable in their quantum version. One has to prove if that is the case depending on the model we are working on. 

Many of the techniques developed to work with classical integrable models, however, like the Yang-Baxter equation and transfer matrices, for example, can be generalized to the quantum case. 

Integrability has very interesting consequences in scattering of particles in (1+1) quantum field theories, for example. The infinity number of conserved charges is responsible for the factorized scattering, where a scattering process with $ n $-particles, can be factorized in a sequence of $ 2\rightarrow 2 $ particles scattering \cite{ZAMOLODCHIKOV1979253}. The two-body S-matrix in such models is related to the R-matrix and has to satisfy the quantum Yang-Baxter equation (qYBE).

The existence of a quantum R-matrix, solution of the quantum Yang-Baxter equation allows for the systematic  construction of the conserved charges, and techniques like Bethe ansatz allow most of the time for the solution of such models.

One very interesting application is in spin chains, which are discrete quantum spin systems which have applications ranging from $ \mathcal{N}=4 $ Super Yang-Mills ($ \mathcal{N}=4 $ SYM) theory to condensed matter. In these lecture notes we discuss a famous example of a spin chain, the XXZ model, which can be understood as a toy model to study magnetism. With this example it is possible to introduce all the important concepts and techniques we need.

\paragraph{Outline of these lecture notes:} In Section \ref{sec:notation} we present a short introduction to Poisson brackets and tensor products. Section \ref{sec:classical} is dedicated to classical integrability, starting with Liouville integrability, Lax pair and Yang-Baxter equation for systems with finite number of degrees of freedom, followed by Lax connection and Yang-Baxter for classical field theory. The next section is dedicated to quantum integrability, more specifically to the construction of the XXZ periodic spin chain and solving it by using the algebraic Bethe ansatz. At the end of this section we shortly discuss the interpretation of the R-matrix as an S-matrix in a (1+1) quantum field theory. In Section \ref{sec:new} we mention some important areas where integrability plays an important role nowadays and provide some references and reviews on these topics. In Appendix \ref{app:proof} we provide a proof for equation \eqref{eq:PoissonL1L2}, while  Appendices \ref{app:Lax} and \ref{app:newmodels} are dedicated to introduce integrable hierarchies and to explain a systematic method to find new solutions of the qYBE, respectively. Along these lecture notes we included many constructive exercises that we believe can help in the understanding of the concepts and techniques.

\section{Some notation and basic properties}\label{sec:notation}

\subsection{Poisson bracket}\label{subsec:Poissonbracket}

Consider a $ 2n $-dimensional phase space with canonical coordinates $ \{q_1,\,q_2,\,...,\,q_n\} $ and $ \{p_1,\,p_2,$ $\,...,\,p_n\} $. For two functions $ F(\{q_i\},\{p_i\}) $ and $ G(\{q_i\},\{p_i\}) $ in this space we can define a \textbf{Poisson bracket} as

\begin{equation}
\left\{F,G\right\}=\sum_{i=1}^{n}\left(\frac{\partial F}{\partial q_i}\frac{\partial G}{\partial p_i}-\frac{\partial F}{\partial p_i}\frac{\partial G}{\partial q_i}\right).
\label{eq:PoissonBracket}
\end{equation}
With this definition it is easy to see that

\begin{align}
& \{q_j,p_k\}=\delta_{j,k},\\
& \{q_j,q_k\}=0,\\
& \{p_j,p_k\}=0.
\end{align}

The Poisson bracket satisfies many important properties

\begin{enumerate}
	\item $ \{F,F\}=0 $;
	\item $ \{G,F\}=-\{F,G\} $;
	\item $ \{F,G+H\}=\{F,G\}+\{F,H\} $;
	\item $ \{FG,H\} =F\{G,H\}+\{F,H\}G$;
	\item $ \{F,GH\} =G\{F,H\}+\{F,G\}H $; and finally the Jacobi identity:
	\item $ \{F,\{G,H\}\} +\{H,\{F,G\}\}+\{G,\{H,F\}\}=0$.
\end{enumerate}

\begin{tcolorbox}

\paragraph{Exercise 1:} Check the properties (1)-(6).
	
\end{tcolorbox}

\subsection{Kronecker product}\label{subsec:Kroneckerproduct}

Both in classical and in quantum integrability \textbf{Kronecker product} (tensor product\footnote{When doing calculations on Mathematica, use the command KroneckerProduct instead of TensorProduct. Although they both give the same result, TensorProduct gives the result in a format that is not useful for the calculations we need to perform.}) is very important. It is a matrix operation that provides a way to deal with systems with more than one vector space $ V $. We dedicate this section to explain it and show some of its properties in a very practical way.

Let us start by defining the tensor product of two vectors. Consider two vector spaces $  U $ ($ d_1- $dimensional ) and $ V $ ($ d_2 $-dimensional), and then consider a vector in each of these vector spaces, $ |u\rangle\in U $ and $ |v\rangle\in V $

\begin{equation}
|u\rangle=\begin{pmatrix}
u_1\\
u_2\\
\vdots\\
u_{d_1}
\end{pmatrix}\quad \text{and} \quad |v\rangle=\begin{pmatrix}
v_1\\
v_2\\
\vdots\\
v_{d_2}
\end{pmatrix}.
\end{equation}
Their tensor product is given by placing $ |v\rangle $ inside $ |u\rangle $ in the following way

\begin{equation}
|u\rangle\otimes |v\rangle=\begin{pmatrix}
u_1\, |v\rangle\\
u_2\, |v\rangle\\
\vdots\\
u_{d_1}\, |v\rangle
\end{pmatrix},
\end{equation}
where the product $ u_i |v\rangle $ is the usual product between a scalar and a vector. Let us consider $  d_1=d_2=2 $ for example: 

\begin{equation}
|u\rangle\otimes |v\rangle=\begin{pmatrix}
u_1 \, |v\rangle\\
u_2 \, |v\rangle
\end{pmatrix}=\begin{pmatrix}
u_1\begin{pmatrix}
v_1\\
v_2
\end{pmatrix}\\
u_2 \begin{pmatrix}
v_1\\
v_2
\end{pmatrix}
\end{pmatrix}=\begin{pmatrix}
u_1 \,v_1\\
u_1 \,v_2\\
u_2 \,v_1\\
u_2 \,v_2 
\end{pmatrix}.
\end{equation}

Now, let us consider instead two $ d\times d $ matrices $ A $ and $ B $ in End$ (V\otimes V) $

\begin{equation}
A=\begin{pmatrix}
a_{1,1} & a_{1,2} & \ldots & a_{1,d}\\
a_{2,1} & a_{2,2} & \ldots & a_{2,d}\\
\vdots & \vdots & \ddots & \vdots\\
a_{d,1} & a_{d,2}& \ldots & a_{d,d}
\end{pmatrix} \quad \text{and} \quad B=\begin{pmatrix}
b_{1,1} & b_{1,2} & \ldots & b_{1,d}\\
b_{2,1} & b_{2,2} & \ldots & b_{2,d}\\
\vdots & \vdots & \ddots & \vdots\\
b_{d,1} & b_{d,2}& \ldots & b_{d,d}
\end{pmatrix} .
\end{equation}
The Kronecker product of $ A $ and $ B $, $ A \otimes B $, is obtained by placing $ B $ inside $ A $ in the following way

\begin{equation}
A\otimes B=\begin{pmatrix}
a_{1,1}B & a_{1,2}B & \ldots & a_{1,d}B\\
a_{2,1}B & a_{2,2}B & \ldots & a_{2,d}B\\
\vdots & \vdots & \ddots & \vdots\\
a_{d,1}B & a_{d,2}B& \ldots & a_{d,d}B
\end{pmatrix}
\end{equation}
where $ a_{i,j}B $ correspond to the usual matrix product between a scalar and a matrix, or a matrix and a matrix, depending if $ a_{i,j} $ is a scalar or a matrix, respectively. 

For example, for a space $ V $ of dimension two with 

\begin{equation}
A=\begin{pmatrix}
a_{1,1} & a_{1,2}\\
a_{2,1} & a_{2,2}
\end{pmatrix}
\quad \text{and}\quad B=\begin{pmatrix}
b_{1,1} & b_{1,2}\\
b_{2,1} & b_{2,2}\end{pmatrix}
\end{equation}
the Kronecker product gives us 
\begin{align}
A\otimes B & =
\begin{pmatrix}
a_{1,1}B & a_{1,2}B\\
a_{2,1}B & a_{2,2}B
\end{pmatrix}\nonumber\\
& =\begin{pmatrix}
a_{1,1}\begin{pmatrix}
b_{1,1} & b_{1,2}\\
b_{2,1} & b_{2,2}
\end{pmatrix} & a_{1,2}\begin{pmatrix}
b_{1,1} & b_{1,2}\\
b_{2,1} & b_{2,2}
\end{pmatrix}\\
a_{2,1}\begin{pmatrix}
b_{1,1} & b_{1,2}\\
b_{2,1} & b_{2,2}
\end{pmatrix} & a_{2,2}\begin{pmatrix}
b_{1,1} & b_{1,2}\\
b_{2,1} & b_{2,2}
\end{pmatrix}
\end{pmatrix}\nonumber\\
&=\begin{pmatrix}
a_{1,1}b_{1,1} & a_{1,1} b_{1,2} & a_{1,2}b_{1,1} &a_{1,2}b_{1,2}\\
a_{1,1}b_{2,1} & a_{1,1} b_{2,2} & a_{1,2}b_{2,1} &a_{1,2}b_{2,2}\\
a_{2,1}b_{1,1} & a_{2,1} b_{1,2} & a_{2,2}b_{1,1} &a_{2,2}b_{1,2}\\
a_{2,1}b_{2,1} & a_{2,1} b_{2,2} & a_{2,2}b_{2,1} &a_{2,2}b_{2,2}
\end{pmatrix}.
\label{KPdefinition}
\end{align}

When describing a system with $ N $ sites $ V\otimes V\otimes ...\otimes V $, for an $ A\in \text{End}(V) $ the notation $ A_i $ means acting non-trivially only on the $ i $-th site, i.e.

\begin{equation}
A_i=\overbrace{\bI\otimes\bI\otimes...\otimes \bI\otimes}^{(i-1)-\text{sites}} \underbrace{A}_{i \text{-th site}}\otimes  \overbrace{\bI\otimes ...\otimes \bI}^{(N-i)-\text{sites}}.
\end{equation}
So, for example, if we are describing a system with $ N=2 $ and $ V=\mathbb{C}^2 $
\begin{align}
&A_1 =A\otimes \mathbb{I}= \begin{pmatrix}
a_{1,1} & 0 & a_{1,2} & 0\\
0 & a_{1,1} & 0 & a_{1,2}\\
a_{2,1} & 0 & a_{2,2} & 0\\
0 & a_{2,1} & 0 & a_{2,2}
\end{pmatrix},\\
&A_2 =\mathbb{I}\otimes A= \begin{pmatrix}
a_{1,1} & a_{1,2} & 0 & 0\\
a_{2,1} & a_{2,2} & 0 & 0\\
0 & 0 & a_{1,1} & a_{1,2}\\
0 & 0 & a_{2,1} & a_{2,2}
\end{pmatrix}.
\end{align}

Now, consider an operator $ B\in \text{End}(V\otimes V) $,  for $ N $ sites, we can define $ B_{i,j} $ as acting non-trivially only on the spaces $ i $ and $ j $. For $ N=3 $, for example 
\begin{align}
&B_{1,2}=B\otimes \bI,\\
&B_{2,3}=\bI\otimes B
\end{align}
i.e., $ B_{1,2} $ acts non-trivially in the first two spaces and as an identity on the third, while $ B_{2,3} $ acts trivially on the first space and non-trivially on the last two spaces.
And how about $ B_{1,3} $?

In order to construct $ B_{1,3} $ we have to introduce another concept: the permutation operator $ P $.
The permutation operator is defined as the object that permutes two vectors:

\begin{equation}
P |a\rangle \otimes |b\rangle=|b\rangle \otimes |a\rangle .
\end{equation}

\begin{tcolorbox}
	
	\paragraph{Exercise 2:} Prove that $ P^2=\mathbb{I} $.
	
\end{tcolorbox}
So, $ P_{ij} $ switches vectors in positions $ i $ and $ j $, for example

\begin{equation}
P_{1,2}|a_1\rangle \otimes |a_2\rangle\otimes |a_3\rangle \otimes.... |a_N\rangle=|a_2\rangle \otimes |a_1\rangle\otimes |a_3\rangle \otimes....\otimes |a_N\rangle
\end{equation}
while 
\begin{equation}
P_{2,3}|a_1\rangle \otimes |a_2\rangle\otimes |a_3\rangle \otimes.... |a_N\rangle=|a_1\rangle \otimes |a_3\rangle\otimes |a_2\rangle \otimes....\otimes |a_N\rangle
\end{equation}
etc.

The same happens for matrices, $ P_{ij} $ basically switches positions $ i \Leftrightarrow j $ so

\begin{equation}
P_{i,j}B_{i,j}P_{i,j}=B_{j,i},
\end{equation}
and
\begin{align}
B_{i,j}& =P_{k,j}B_{i,k}P_{k,j},\nonumber\\
&=P_{i,k}B_{k,j}P_{i,k}.
\label{eq:PAPproperty}
\end{align}
Now, using equation \eqref{eq:PAPproperty} we can construct $ B_{1,3} $ in two ways

\begin{equation}
B_{1,3}=P_{1,2}B_{2,3}P_{1,2} \quad \text{and}\quad B_{1,3}=P_{2,3}B_{1,2}P_{2,3}.
\end{equation} 

So, when considering a system with many sites, there are many equivalent ways to construct $ B_{i,j} $, given that $ k $ can assume many different values in equation \eqref{eq:PAPproperty}.

\begin{tcolorbox}

	\paragraph{Exercise 3:} Consider an operator $ B\in \text{End}(V\otimes V) $ and the permutation operator $ P\in \text{End}(V\otimes V) $. Using the properties discussed above, for $ N=5 $, construct $ P_{2,4} $ and $ B_{2,5} $ using only operators acting on consecutive sites.\\

\end{tcolorbox}

\begin{tcolorbox}
	
	\paragraph{Exercise 4:} It is very useful to notice that $ \tr_{i}P_{i,j}=I_j $. Prove this using
	
	\begin{equation}
	P=\sum_{i,j=1}^{2}e_{ij}\otimes e_{ji}
	\label{eq:Pexercise}
	\end{equation}
	where $ (e_{ij})_{\alpha\beta}=\delta_{i\alpha}\delta_{j\beta}  $, i.e., 
	\begin{equation}
	e_{11}=\begin{pmatrix}
	1 & 0\\
	0 & 0
	\end{pmatrix}, \quad e_{12}=\begin{pmatrix}
	0 & 1\\
	0 & 0
	\end{pmatrix}, \quad \text{etc.}
	\end{equation}
	
\end{tcolorbox}

\begin{tcolorbox}

	\paragraph{Exercise 5:} Prove
	\begin{equation}
	(P_{1,2}P_{2,3}...P_{N-1,N})^{-1}=P_{N-1,N}...P_{2,3}P_{1,2}.
	\label{eq:InverseofP}
	\end{equation}
	
\end{tcolorbox}

It is useful to notice that for $V=\mathbb{C}^2 $ we can also write $ P $ as

\begin{align}
P&=\begin{pmatrix}
1 & 0 & 0 & 0\\
0 & 0 & 1 & 0\\
0 & 1 & 0 & 0\\
0 & 0 & 0 & 1
\end{pmatrix}=\begin{pmatrix}
\frac{1+\sigma^z}{2} & \sigma^-\\
\sigma^+&\frac{1-\sigma^z}{2} 
\end{pmatrix},\\
&=\frac{1}{2}\left(\sigma_i^x\sigma_{i+1}^x+\sigma_i^y\sigma_{i+1}^y+\sigma_i^z\sigma_{i+1}^z+I_i\,I_{i+1}\right)
\end{align}
where $ \sigma^\pm=\frac{1}{2}\left(\sigma^x\pm i \sigma^y\right) $ and $ \sigma^{x,y,z} $ are the Pauli matrices. 

Some extra properties that play a role on these lectures are 
\begin{enumerate}
	\item $ (A+B)\otimes C=A\otimes C+B\otimes C $;
	\item $ A\otimes (B+C)=A\otimes B+A\otimes C $;
	\item $ (cA)\otimes B=A\otimes (cB)=c(A\otimes B) $;\
	
where $ A,B,C,D $ are matrices and $ c $ is a scalar function.

Also, if the dimensions of the matrices are such that the product $ AC $ and $ BD $ are well defined then
	\item $ (A\otimes B) (C\otimes D)=(AC)\otimes (BD)$.
\end{enumerate}

\begin{tcolorbox}
	
	\paragraph{Exercise 6:} For $ A,B \in \text{End}(V\otimes V) $, understand why 
	
	\begin{equation}
	\left[A_{12},B_{34}\right]=0
	\label{eq:diffspaces}
	\end{equation}
	is true for any two matrices $ A $ and $ B $ of same dimension.
	
\end{tcolorbox}

Something that will appear often in the lectures, especially in the classical part, is the Poisson bracket of matrices. Now that we defined Kronecker product there is a simple way to introduce them. 

Consider a matrix $ A \in  \text{End}(V)  $

\begin{equation}
A=\sum_{i,j}a_{ij}e_{ij},
\end{equation} 
so that 

\begin{align}
& A_1=\sum_{i,j}a_{ij}(e_{ij}\otimes \mathbb{I}),\\
& A_2=\sum_{k,l}a_{kl}(\mathbb{I}\otimes e_{kl})
\end{align}
where again $ (e_{ij})_{\alpha\beta}=\delta_{i\alpha}\delta_{j\beta}  $.

The Poisson bracket of $ A_1 $ and $ A_2 $ is defined as

\begin{equation}
\left\{A_1,A_2\right\}=\sum_{i,j,k,l}\left\{a_{ij},a_{kl}\right\}(e_{ij}\otimes e_{kl}).
\end{equation}

\begin{tcolorbox}
	
	\paragraph{Exercise 7:} Assume the properties (1)-(6) from Section \ref{subsec:Poissonbracket} are valid for $ F $ and $ G $ being matrices whose elements are functions of $ \{q_i\} $ and $ \{p_i\} $. Prove that if $ B $ is an invertible matrix then
	
	\begin{equation}
	\left\{A_1,B_2^{-1}\right\}=-B_2^{-1}\left\{A_1,B_2\right\}B_2^{-1}.
	\label{eq:Poissoninverse}
	\end{equation}

\end{tcolorbox}

\section{Classical integrability}\label{sec:classical}

Now, with all the prerequisites let us start the discussion of integrability. Some of the discussions in this section are highly based in the amazing book \cite{babelon_bernard_talon_2003}. 

Consider a function $ F(\{q_j\},\,\{p_j\} $)  in a $ 2n $-dimensional phase space described by the conjugated variables $ \{q_j\} $ and $ \{p_j\} $

\begin{equation}
\dot{F}\equiv \frac{dF}{dt}=\sum_{i=1}^{n}\left(\frac{\partial F}{\partial q_i}\dot{q}_i+\frac{\partial F}{\partial p_i}\dot{p}_i\right).
\label{eq:differentialF}
\end{equation}
We know, however, that the Hamilton equations are given by

\begin{equation}
\dot{q}_i=\frac{\partial H}{\partial p_i}, \quad \text{and} \quad \dot{p}_i=-\frac{\partial H}{\partial q_i}.
\label{eq:Hamiltonequations}
\end{equation}
Substituting them in the equation \eqref{eq:differentialF} we obtain

\begin{equation}
\dot{F}=\sum_{i=1}^{n}\left(\frac{\partial F}{\partial q_i}\frac{\partial H}{\partial p_i}-\frac{\partial F}{\partial p_i}\frac{\partial H}{\partial q_i}\right)
\end{equation}
where the rhs can be recognized as the Poisson bracket of $ F $ and $ H $ 
\begin{equation}
\dot{F}=\{F,H\}.
\label{eq:evolutionofF}
\end{equation}
So, any function of $ \left\{p_i\right\} $ and $ \left\{q_i\right\} $ in the phase space will have its evolution described by the equation \eqref{eq:evolutionofF}.

From this we can immediately say that any function $ F $ satisfying $ \{F,H\} =0$ is conserved. 
The Hamiltonian itself is the most obvious example 
\begin{equation}
\dot{H}=\{H,H\}=0\Rightarrow H=E=\text{constant}.
\end{equation}

\subsection{Liouville integrability}\label{sec:Liouville}

A system in the $ 2n $-dimensional phase space described above is \textbf{Liouville integrable} if it has $ n $ independent conserved quantities in involution, i.e.
\begin{equation}
\left\{F_i,F_j\right\}=0, \quad i,j=1,...,n.
\label{eq:Liouvilleintegrability}
\end{equation}
One of these is the Hamiltonian $ H $.

The importance of such systems comes from the fact that they are completely solvable, as a consequence of Liouville's  theorem:

\paragraph{Liouville Theorem:} the equations of motion of a Liouville integrable system can be solved by quadratures.

Due to this theorem we know that for a Liouville integrable system there always exists a canonical transformation\footnote{A canonical transformation is a change in the canonical variables in such a way that the Hamilton equations form is preserved.}

\begin{equation}
(p_i,q_i)\rightarrow (F_i,\varphi_i),
\end{equation}
where one of the new variables coincides with the conserved quantity $ F_i $, whose equations of motion can be described by 

\begin{align}
& \dot{F_i}=\left\{H,F_i\right\}=0,\label{eq:Fdot}\\
& \dot{\varphi}_i=\left\{H,\varphi_i\right\}=\frac{\partial H}{\partial F_i},=\Omega_i\label{eq:varphidot}
\end{align}
with the following solution
\begin{equation}
F_i(t)=\alpha \quad \text{and} \quad \varphi_i(t)=\Omega_i\, t+\varphi_i(0).
\end{equation}

\paragraph{Example:} 1d Harmonic-Oscillator

The  Hamiltonian for the 1d classical harmonic oscillator, for mass $ m=1 $, is given by

\begin{equation}
H=\frac{p^2}{2}+\frac{\omega^2q^2}{2}=E, \quad \omega\in\mathbb{R}.
\end{equation}
We can rewrite this in terms of the new variables 
\begin{align}
&p=\rho \cos\varphi,\label{eq:newvariablep}\\
&q=\frac{\rho}{\omega}\sin\varphi, \label{eq:newvariableq}
\end{align}
and obtain 
\begin{align}
H&=\frac{\rho^2}{2}\cos^2\varphi+\frac{\omega^2}{2}\frac{\rho^2}{\omega^2}\sin^2\varphi,\nonumber\\
&=\frac{\rho^2}{2}=E
\end{align}
which means that $\rho=\sqrt{2E}$. 

\begin{tcolorbox}
	\paragraph{Exercise 8:} Check that equations \eqref{eq:Fdot} and \eqref{eq:varphidot} are satisfied with $ \{\rho,\varphi\}=\omega/\rho $, $ F_i=\rho $ and $ \Omega_i=\omega $ and therefore
	
	\begin{equation}
	\varphi(t)=\omega \,t+\varphi(0). 
	\end{equation}
	
\end{tcolorbox}

Liouville's theorem is very powerful and plays an important role in classical integrability. But there is another formalism that possesses many advantages in comparison: the Lax pair and the Classical Yang-Baxter equation. The main advantage of this procedure is that one can naturally generalize it to describe (1+1) integrable field theories by using the so-called Lax connection.
That is what we will discuss in sections \ref{subsec:LaxCYBE} - \ref{subsec:classicalmonodromy}.

\subsection{Lax pair and Classical Yang-Baxter equation}\label{subsec:LaxCYBE}

As mentioned above, for some class of models, there is a more systematic way to work with classical integrable systems. In order to do this we need the so called Lax pair. There are two cases: the ``constant" Lax pair and the Lax pair depending on a spectral parameter $ z $.

\subsubsection{``Constant" Lax pair}\label{subsubsec:constantLax}

We learned in the previous section that Liouville integrability means that

\begin{enumerate}
	\item we have a tower of conserved charges (one charge for each degree of freedom);
	\item these charges are all in involution.
\end{enumerate}

We will consider these two points separately in order to construct the Lax pair $ (L,M) $ which generates these conserved charges. 

Let us start by addressing the first point. Notice that it is possible to write the Hamilton equations \eqref{eq:Hamiltonequations} in a matrix form given by

\begin{equation}
\dot{L}=[M,L]
\label{eq:dotL=ML}
\end{equation}
where $ M $ and $L $ are called \textbf{Lax pair}, and $ [\boldsymbol{\cdot},\boldsymbol{\cdot}] $ is the commutator. The exact form of $ L $ does not matter at this stage. Conditions on it will be however necessary in order to assure that the charges are in involution. 

The conserved charges are given by 

\begin{equation}
Q_n=\tr L^n
\label{eq:conservedQclassical}
\end{equation}
as we check below.

In order to check that it is indeed the case, we can take the derivative with respect to $ t $ and obtain

\begin{align}
\dot{Q_n}&=\tr \left(\dot{L}\,L^{n-1}\right)+\tr \left(L\,\dot{L}\,L^{n-2}\right)+...+\tr \left(L^{n-2}\,\dot{L}\,L\right)+\tr \left(L^{n-1}\,\dot{L}\right),\nonumber\\
&=n\,\tr\left( \dot{L}\,L^{n-1}\right),\nonumber\\
&=n\,\tr\left(\left[M,L\right]L^{n-1}\right),\nonumber\\
&=n\,\tr\left(M\,L\,L^{n-1}-L\,M\,L^{n-1}\right),\nonumber\\
&=n\,\tr\left(M\,L^{n}\right)-n\,\tr\left(M\,L^{n}\right),\nonumber\\
&=0,
\end{align}
where we use multiple times the cyclicity  of the trace, and also equation \eqref{eq:dotL=ML}.

\paragraph{Example:} Let us return to the harmonic oscillator example. Its Lax pair is given by
	\begin{equation}
	L=\frac{1}{2}\begin{pmatrix}
	p & \omega \,q\\
	\omega\,q& -p
	\end{pmatrix} \quad \text{and} \quad M=\frac{1}{2}\begin{pmatrix}
	0 & -\omega\\
	\omega & 0
	\end{pmatrix}.
	\end{equation} 
	Substituting it in the equation \eqref{eq:dotL=ML} we obtain the usual equations of motion. 
	
	Computing $ L^2 $ we find 
	
	\begin{equation}
	L^2=\frac{1}{4}\begin{pmatrix}
	p^2+\omega^2q^2 & 0\\
	0&p^2+\omega^2q^2
	\end{pmatrix}.
	\end{equation}
	So, $H=\tr L^2$.
 
\begin{tcolorbox}

	\paragraph{Exercise 9:} Consider the \textbf{Calogero oscillator model} (see \cite{Arutyunov2019}) which is described by the following Hamiltonian 
	
	\begin{equation}
	H=\frac{p^2}{2}+\frac{\omega^2q^2}{2}+\frac{\nu^2}{2q^2}.
	\end{equation}
	
	This model is integrable and can be described by the following Lax pair
	
	\begin{equation}
	L=\frac{1}{2}\begin{pmatrix}
	p & \omega \,q-\frac{\nu}{q}\\
	\omega\,q-\frac{\nu}{q} & -p
	\end{pmatrix} \quad \text{and} \quad M=\frac{1}{2}\begin{pmatrix}
	0 & -\omega-\frac{\nu}{q^2}\\
	\omega+\frac{\nu}{q^2} & 0
	\end{pmatrix}.
	\end{equation}
	
	This model becomes the harmonic oscillator for $ \nu\rightarrow 0 $ and the rational Calogero model for $ \omega\rightarrow 0 $.\\
	\textbf{a)} Compute the equations of motion using the Hamilton equations from the Hamiltonian and using the Lax pair. Check that the results match.\\
	\textbf{b)} Check that 
	\begin{equation}
	H=\tr L^2+ \kappa
	\end{equation}
	and find the constant $ \kappa $.
\end{tcolorbox}

Also, the equation \eqref{eq:dotL=ML} has a simple solution

\begin{equation}
L(t)=g(t)L(0)g(t)^{-1}, \quad M(t)=\frac{dg(t)}{dt}g(t)^{-1}.
\end{equation}
This is important because it means that any function of $ L $ that is invariant by conjugation is a constant of motion. The equation \eqref{eq:dotL=ML} is called \textbf{isospectral} because the spectrum of the Lax matrix is preserved by the time evolution.

Now, we need to address the question of whether the conserved quantities are in involution or not.

Suppose that $ L $ can be diagonalized by a matrix $ U $

\begin{equation}
L=U \Lambda U^{-1}.
\label{eq:Ldiagonalized}
\end{equation}
This means that elements $ \Lambda_{ii} $ will be conserved. So, for the charges to be in involution we need

\begin{equation}
\left\{\Lambda_1,\Lambda_2 \right\}=0.
\label{eq:commutationLambdas}
\end{equation}

By computing $ \left\{L_1,L_2\right\} $, using equations \eqref{eq:Ldiagonalized} and \eqref{eq:commutationLambdas} one obtains

\begin{equation}
\left\{L_1,L_2\right\}=\left[r_{12},L_1\right]-\left[r_{21},L_2\right]
\label{eq:PoissonL1L2}
\end{equation}
where $ r_{12} $ is called r-matrix and it depends on $ U_{1,2} $ and $ z_{1,2} $. The proof is a bit long and we placed it in the appendix \ref{app:proof}. Notice that all lower indices in the matrices are in the Kronecker product sense introduced in section \eqref{subsec:Kroneckerproduct}.

We know that the Poisson bracket satisfies the Jacobi identity

\begin{equation}
\left\{L_1,\left\{L_2,L_3\right\}\right\}+\left\{L_3,\left\{L_1,L_2\right\}\right\}+\left\{L_2,\left\{L_3,L_1\right\}\right\}=0.
\end{equation}
Substituting equation \eqref{eq:PoissonL1L2} in the Jacobi identity a few times and simplifying it, we obtain the following equation
\begin{align}
&\left[L_1,\left\{L_2,r_{13}\right\}-\left\{L_3,r_{12}\right\}+\left[r_{12},r_{13}+r_{23}\right]+\left[r_{32},r_{13}\right]\right]+\nonumber\\
&\left[L_2,\left\{L_3,r_{21}\right\}-\left\{L_1,r_{23}\right\}+\left[r_{23},r_{21}+r_{31}\right]+\left[r_{13},r_{21}\right]\right]+\nonumber\\
&\left[L_3,\left\{L_1,r_{32}\right\}-\left\{L_2,r_{31}\right\}+\left[r_{31},r_{32}+r_{12}\right]+\left[r_{21},r_{32}\right]\right]=0
\label{eq:JacobiYBE}
\end{align}
which is composed of three terms with Poisson brackets and three terms with only commutators.

If $ r_{ij} $ does not depend on the dynamical variables, all the Poisson brackets become zero in equation \eqref{eq:JacobiYBE}. Also, if $ r_{ij} $ satisfies $ r_{ij}=-r_{ji} $ we obtain the so called \textbf{classical Yang-Baxter (cYBE) equation}

\begin{equation}
\left[r_{12},r_{13}+r_{23}\right]+\left[r_{13},r_{23}\right]=0.
\label{eq:cYBE}
\end{equation}

\
\begin{tcolorbox}
	
	\paragraph{Exercise 10:} The matrix $ U $ that diagonalizes the Lax matrix $ L $ of the Harmonic oscillator is given, in terms of the variables $ \rho $ and $ \varphi $ introduced in equations \eqref{eq:newvariablep}-\eqref{eq:newvariableq}, by
	
	\begin{equation}
	U=\begin{pmatrix}
	\cos\left(\frac{\varphi}{2}\right) & \sin\left(\frac{\varphi}{2}\right)\\
	\sin\left(\frac{\varphi}{2}\right) & -\cos\left(\frac{\varphi}{2}\right)
	\end{pmatrix}.
	\end{equation}
	From the proof on Appendix \ref{app:proof} we learned that 
	
	\begin{equation}
	r_{12}=U_2\left\{U_1,\Lambda_2\right\}U_1^{-1}U_2^{-1}+\frac{1}{2}\left[\left\{U_1,U_2 \right\}U_1^{-1}U_2^{-1},L_2\right].
	\end{equation}
	With this in mind, compute the r-matrix $ r_{12} $ for the Harmonic oscillator. Does it satisfy the classical Yang-Baxter in the form \eqref{eq:cYBE}?
	
\end{tcolorbox}

\subsubsection{Lax pair depending on a spectral parameter}\label{subsubsec:nonconstantLax}

There are some models, however, that cannot be described by the formalism discussed above. Actually, the most interesting cases appear when we add a new parameter to this description, the \textbf{spectral parameter} $ z \in \mathbb{C} $. This parameter, in principle is not physical, so the calculations have to work for any value of it. With this, the construction of a whole tower of conserved charges can be obtained by an expansion in this parameter.

In this section, we discuss such generalization, but without repeating all the explanations and proofs, focusing however on highlighting the most important differences between both cases. 

The equation \eqref{eq:dotL=ML} is substituted by

\begin{equation}
\dot{L}(z)=\left[M(z),L(z)\right],
\label{eq:dotL=MLlambda}
\end{equation}
so, the evolution is given by

\begin{equation}
L(z)\Psi =\Lambda(z)\Psi
\end{equation}
while \eqref{eq:PoissonL1L2} becomes

\begin{equation}
\left\{L_1(z_1),L_2(z_2)\right\}=\left[r_{12}(z_1-z_2),L_1(z_1)\right]-\left[r_{21}(z_2-z_1),L_2(z_2)\right],
\label{eq:PoissonL1L2lambda}
\end{equation}
and the cYBE becomes

\begin{equation}
\left[r_{12}(z_1-z_2),r_{13}(z_1-z_3)+r_{23}(z_2-z_3)\right]+\left[r_{13}(z_1-z_3),r_{23}(z_2-z_3)\right]=0
\label{eq:cYBElambda}
\end{equation}
which can be rewritten as

\begin{equation}
\left[r_{12}(z_1-z_2),r_{13}(z_1)\right]+\left[r_{12}(z_1-z_2),r_{23}(z_2)\right]+\left[r_{13}(z_1),r_{23}(z_2)\right]=0
\label{eq:cYBElambda2}
\end{equation}
because $ z_3 $ can be absorbed in $ z_1 $ and $ z_2 $. So, equation \eqref{eq:cYBElambda2} is the \textbf{classical Yang-Baxter equation} for a spectral parameter dependent r-matrix.

Notice that in order to obtain \eqref{eq:cYBElambda2} using the Jacobi identity one has to assume

\begin{equation}
r_{12}(z)=-r_{21}(-z).
\label{eq:r12=-r21}
\end{equation} 
We also assumed that the r-matrix depends only on the difference of spectral parameters, i.e., $ r_{i,j}(z_1,z_2) =r_{i,j}(z_1-z_2)$.

Notice however, that \eqref{eq:cYBElambda2} is not the most general form of cYBE. We could have in the rhs a Casimir element, i.e., an element that commutes with all the $ L_i $'s, and the spectral parameter dependent version of \eqref{eq:JacobiYBE} would still be satisfied. In such case, the equation is called \textbf{modified classical Yang-Baxter equation} \cite{babelon_bernard_talon_2003} and it plays an important role in integrable sigma models as it is discussed in \cite{Hoare:2021dix}. 

Actually, there is a classification of the r-matrices that satisfy equation \eqref{eq:r12=-r21}. The classification is due to Belavin and Drinfeld and it says that all the poles of the r-matrix are simple and if they form a:

\begin{itemize}
	\item 0-dimensional lattice, then the r-matrices are called rational (because they depend on rational functions only);
	\item 1-dimensional lattice, then the r-matrices are called trigonometric(because they depend on trigonometric functions only);
	\item 2-dimensional lattice, then the r-matrices are called elliptic(because they depend on elliptic functions only);
\end{itemize}
So, there is a direct relation between the number of poles and the form of the r-matrices. This will appear in a very interesting way in the 4-dimensional Chern-Simons theory in Sylvain Lacroix's lectures.

From now on, in this lectures our examples are focused on the trigonometric r-matrices. This is because the example we focus on the quantum part is the XXZ spin chain, which corresponds to the quantization of a trigonometric classical r-matrix. 

The complete classification for trigonometric cases can be found in \cite{Belavin1982, Jimbo:1985ua}. These models are called generalized Toda systems. They are written in a simple way in the paper \cite{Jimbo:1985ua} by Jimbo where he generalizes them to the quantum case.  We will discuss more about this in section \ref{sec:quantum}. But, all the r-matrices for the affine Lie algebras are of trigonometric form and depend on the spectral parameter $ z $. 

Let us show now, some examples of these r-matrices:

\paragraph{Example: $ A_1^{(1)} \, (\widehat{sl}(2)) $ r-matrix}

\begin{equation}
r(z)=\begin{pmatrix}
\coth\,z & 0 & 0 & 0\\
0 & 0 & \text{csch}\,z & 0\\
0 & \text{csch}\,z & 0 & 0\\
0 & 0 & 0 & \coth \,z
\end{pmatrix}
\label{eq:C11classical}
\end{equation}
This is actually the classical limit of the R-matrix for the spin 1/2 XXZ spin chain, that will be discussed in section \ref{sec:quantum}.

The next example, is the classical version of the spin 1 XXZ spin chain:

\paragraph{Example: $ A_2^{(1)} \, (\widehat{sl}(3)) $ r-matrix}

\begin{equation}
r(z)=\left(
\begin{array}{ccccccccc}
-2\,f(z ) & 0 & 0 & 0 & 0 & 0 & 0 & 0 & 0 \\
0 & f(z ) & 0 & e^{-\frac{z}{2} }g(z ) & 0 & 0 & 0 & 0 & 0 \\
0 & 0 & f(z ) & 0 & 0 & 0 & e^{-\frac{z}{2} }g(z ) & 0 & 0 \\
0 & e^{\frac{z}{2} } g(z ) & 0 & f(z ) & 0 & 0 & 0 & 0 & 0 \\
0 & 0 & 0 & 0 & -2\,f(z ) & 0 & 0 & 0 & 0 \\
0 & 0 & 0 & 0 & 0 & f(z ) & 0 & e^{-\frac{z}{2} }g(z ) & 0 \\
0 & 0 & e^{\frac{z}{2}} g(z ) & 0 & 0 & 0 & f(z ) & 0 & 0 \\
0 & 0 & 0 & 0 & 0 & e^{\frac{z}{2} } g(z ) & 0 & f(z ) & 0 \\
0 & 0 & 0 & 0 & 0 & 0 & 0 & 0 & -2\,f(z ) \\
\end{array}
\right)
\label{eq:A21classical}
\end{equation}
where\footnote{In comparison with \cite{Jimbo:1985ua}) we have $ x\rightarrow e^z $. This is because Jimbo uses YBE in a form depending on $ r_{12}(z_1/z_2) $ instead of $ r_{12}(z_1-z_2) $.}

\begin{equation}
f(z)=\frac{1}{3}\coth\left(\frac{z}{2}\right) \quad \text{and} \quad g(z)=-\text{csch}\left(\frac{z}{2}\right).
\end{equation}

\begin{tcolorbox}
	
	\paragraph{Exercise 11:} For the $ A_1^{(1)} $ and $ A_2^{(2)} $ models presented above\\
	\textbf{a)} Check that the two examples above satisfy $ r_{12}(z)=-r_{21}(-z) $.\\
	\textbf{b)} Check that they satisfy the cYBE \eqref{eq:cYBElambda2}. 
	In  \cite{Jimbo:1985ua} one can find all the generalized Toda systems (classical and quantum), in case the reader would like to work with any of them.
\end{tcolorbox}

\subsection{Lax connection and integrable field theories}\label{subsec:LaxconnectionIFT}

In order to construct the charges in the previous sections we heavily focused on Liouville integrability. When we need to discuss field theories though, this does not make immediate sense, given that we now have an infinite number of degrees of freedom.

We can approach this problem by starting with an auxiliary problem
\begin{align}
&\partial_x\Psi(x,t;z)+ \mathcal{L}_x(x,t;z)\Psi(x,t;z)=0,\\
&\partial_t\Psi(x,t;z)+ \mathcal{L}_t(x,t;z)\Psi(x,t;z)=0
\label{eq:auxiliaryproblem}
\end{align}
and by requiring the compatibility condition (i.e. $ \partial_x\partial_t\Psi= \partial_t\partial_x\Psi $) we obtain the \textbf{flat connection condition} 
 
\begin{equation}
\partial_t\mathcal{L}_x-\partial_x\mathcal{L}_t+\left[\mathcal{L}_t,\mathcal{L}_x\right]=0
\label{eq:zerocurvaturerepresentation}
\end{equation}
also known as the \textbf{zero curvature representation}, which gives us a notion of parallel transport; $ \mathcal{L}_x$ and $\mathcal{L}_t $ are gauge potentials and can be thought as the components $ x $ and $ t $ of a connection called \textbf{Lax connection}.

For a systematic construction of the Lax connection associated with some algebras it is usually convenient to work with Laurent series in $ z $

\begin{equation}
\cL_x=\sum_{i=-n}^{n}\cL_x^{(i)}z^i\quad \text{and}\quad \cL_t=\sum_{i=-m}^{m}\cL_t^{(i)}z^i, \quad m,n\in\mathbb{Z^+},
\end{equation}  
and then use properties of the underlying algebra to find the coefficients $ \cL_x^{(i)} $ and $ \cL_t^{(i)} $.

We discuss below, in the examples and exercises some cases where the Lax connection is known, and in the appendix \ref{app:Lax} we explain how to systematically construct the Lax connection for the equations in two integrable hierarchies called mKdV and AKNS. 

Let us now talk a little about a model that plays an important role in integrability and use it to exemplify the Lax connection.

\paragraph{Example:} Sine-Gordon (S-G) model 

The Sine-Gordon model is probably one of the most well known integrable models. It appears in both classical and quantum integrability, it has two supersymmetric versions and it plays a role in areas that range from condensed matter to string theory. Its equation of motion is a nonlinear differential equation given by

\begin{equation}
\partial_t^2\phi-\partial_x^2\phi=2\sin(2\phi),
\label{eq:SGequation}
\end{equation}
which has solitons as its solutions. 

The Lax for this model is known, and it is given by the following expansion
\begin{align}
\cL_x=\frac{\cL_x^{(-1)}}{z}+\cL_x^{(0)}+\cL_x^{(1)}z\\
\cL_t=\frac{\cL_t^{(-1)}}{z}+\cL_t^{(0)}+\cL_t^{(1)}z
\label{eq:expansionSG}
\end{align}
with coefficients

\begin{equation}
\cL_x^{(-1)}=\frac{i}{2}(e^{-i\,\phi}\sigma^+-e^{i\,\phi}\sigma^-), \quad \cL_x^{(1)}=\left(\cL_x^{(-1)}\right)^*,\quad \cL_x^{(0)}=\frac{i}{2}\partial_t\phi\,\sigma^z,
\end{equation}
\begin{equation}
\cL_t^{(-1)}=-\cL_x^{(-1)}, \quad \cL_t^{(1)}=\cL_x^{(1)},\quad \cL_t^{(0)}=\frac{i}{2}\partial_x\phi\,\sigma^z.
\end{equation}
Substituting the above expressions in the zero curvature equation we are able to generate the Sine-Gordon equation \eqref{eq:SGequation}.

\begin{tcolorbox}
	
	\paragraph{Exercise 12:} Check that substituting the above expansion \eqref{eq:expansionSG} into the zero curvature equation one obtains the Sine-Gordon equation.
	
\end{tcolorbox}

\begin{tcolorbox}
	\paragraph{Exercise 13:} Consider the following expansion \cite{Faddeevbooks}
	
	\begin{equation}
	\cL_x=\cL_x^{(0)}+z \cL_x^{(1)} \quad \text{and} \quad \cL_t=\cL_t^{(0)}+z \cL_t^{(1)}+z^2 \cL_t^{(2)}
	\end{equation}
	with 
	\begin{align}
	& \cL_x^{(0)}=\sqrt{\kappa}(\psi^*\sigma^++\psi\sigma^-), \quad \cL_x^{(1)}=\frac{\sigma^z}{2\,i}, \quad \cL_t^{(1)}=\frac{1}{2}\cL_x^{(0)},\quad \cL_t^{(2)}=\frac{1}{2}\cL_x^{(1)},\\
	& \cL_t^{(0)}=-\frac{i}{2} \kappa |\psi|^2 \sigma^z  -\frac{i}{2} \,\sqrt{\kappa}\left(\partial_x\psi^*\sigma^+-\partial_x\psi \sigma^-\right).
	\end{align}
	Substitute these expansions into the zero curvature equation to find the Non-linear Schroedinger (NLS) equation. 
\end{tcolorbox}

\subsection{The monodromy matrix}\label{subsec:classicalmonodromy}

Continuing with the discussion, let us see how the conserved quantities appear in this context. We will focus on periodic boundary conditions. We assume that the wave function $ \Psi(x,t;z) $ has initial condition $ \Psi(0,0;z)=1 $, and we define a path $ \gamma $ from the origin $ (0,0) $ to the point $ (x,t) $ using the gauge potentials and a path-ordered exponential as

\begin{equation}
\Psi(x,t;z)=\overleftarrow{\text{exp}}\left[-\int_{\gamma} \mathcal{L}_x dx- \mathcal{L}_t dt\right].
\end{equation}

The value of the path-ordered exponential can not depend on the path $ \gamma $ given that the Lax connection satisfies the zero curvature equation.  So, let us consider a path $ \gamma $ given by $ x=[0,2\pi] $ and a fixed time. We can then define the so-called \textbf{monodromy matrix} as the path-ordered exponential

\begin{equation}
T(z)=\overleftarrow{\text{exp}}\left[-\int_0^{2\pi}\cL_x(x;z)dx\right].
\label{eq:monodromyclassical}
\end{equation}

Let us now compute the evolution of $ T(z) $

\begin{align}
\partial_tT(z)&=-\int_0^{2\pi}\left[\left(\overleftarrow{\text{exp}}\int_x^{2\pi}-\cL_x(x^\prime;z)dx^\prime\right)\partial_t\cL_x(x;z)\left(\overleftarrow{\text{exp}}\int_0^{x}-\cL_x(x^\prime;z)dx^\prime\right)\right]dx,\\
&=-\int_0^{2\pi}\left[\left(\overleftarrow{\text{exp}}\int_x^{2\pi}-\cL_x(x^\prime;z)dx^\prime\right)\left(\partial_x\cL_t(x;z)-\left[\cL_t(x;z),\cL_x(x;z)\right]\right)\times\right.\nonumber\\
&\hspace{1cm}\left.\times\left(\overleftarrow{\text{exp}}\int_0^{x}-\cL_x(x^\prime;z)dx^\prime\right)\right]dx\\
&=-\int_0^{2\pi}\partial_x\left[\left(\overleftarrow{\text{exp}}\int_x^{2\pi}-\cL_x(x^\prime;z)dx^\prime\right)\cL_t(x;z)\left(\overleftarrow{\text{exp}}\int_0^{x}-\cL_x(x^\prime;z)dx^\prime\right)\right]dx\\
&=-\left[\left(\overleftarrow{\text{exp}}\int_x^{2\pi}-\cL_x(x^\prime;z)dx^\prime\right)\cL_t(x;z)\left(\overleftarrow{\text{exp}}\int_0^{x}-\cL_x(x^\prime;z)dx^\prime\right)\right]\Big|_{0}^{2\pi}\\
&=-\cL_t(2\pi;z)\overleftarrow{\text{exp}}\left[\int_0^{2\pi}-\cL_x(x;z)dx\right]+\overleftarrow{\text{exp}}\left[\int_0^{2\pi}-\cL_x(x;z)dx\right]\cL_t(0;z)\\
&=-\cL_t(2\pi;z)T(z)+T(z)\cL_t(0;z)
\end{align}
so, considering periodic boundary conditions

\begin{equation}
 \cL_t(0;z)=\cL_t(2\pi,z)
\end{equation}
we find

\begin{equation}
\partial_tT(z)=\left[T(z),\cL_t(0;z)\right].
\label{eq:}
\end{equation}

This is a Lax equation! So, we can interpret $ \cL_t(0;z) $ as $M(z) $ and the monodromy $ T(z) $ as the Lax $ L(z) $. For this reason the quantity

\begin{equation}
t(z)=\tr T(z), 
\end{equation}
which is called transfer matrix, is conserved. In order to guarantee the involution 
\begin{equation}
\{t(z_1),t(z_2)\}=0
\end{equation}
it is enough that the Poisson bracket of the monodromy matrices $ T_1(z_2) $  and $ T_2(z_2) $ satisfies

\begin{equation}
\left\{T_1(z_1),T_2(z_2)\right\}=\left[r_{12}(z_1-z_2)\,,\,T_1(z_1)\,T_2(z_2)\right].
\end{equation}

The $ t(z) $ is called transfer matrix and can be written as an expansion in the spectral parameter

\begin{equation}
t(z)=\sum_{i=1}Q_iz^i, \quad i\in \mathbb{Z}^+
\label{eq:texpandedonz}
\end{equation}
and the coefficients $ Q_i $ are the conserved charges, in involution due to the involution of $ t(z) $ itself.

In order to learn more about Lax formulation, in addition to the already cited references, see also \cite{Dickey,Torrielli:2016ufi,Lacroix:2018njs,Gaiotto:2009hg}.

\section{Quantum integrability}\label{sec:quantum}

In the previous sections we focused on the classical aspects of integrability. Now we intend to discuss their quantization. Many of the objects we introduced there as the r-matrix, the Lax operators, the monodromy and the YBE appear here in some sense as well. The focus now however is on a lattice model. 

We construct all this part focusing on one example: the XXZ periodic spin chain. This model is an important toy model of magnetism and has all the ingredients we need to introduce a quantum integrable model. All the constructions presented here are however very general and can be applied to other models. 
In this section we intend to construct and solve a quantum spin chain using the Algebraic Bethe ansatz \cite{Bethe1931,YangYang1969,Baxter:1982zz,Korepin:1993,Faddeev:1996iy,Sklyanin1979new,Izergin:1984tq,Drinfeld1986quantum}.

\subsection{The XXZ spin chain: the Hamiltonian}\label{subsec:XXZspinchain}

The Hamiltonian $ \mathbb{H} $ for XXZ periodic spin chain is given by 
\begin{align}
& \mathbb{H}=-\frac{J}{2}\sum_{i=1}^{N}H_{i,i+1}=-\frac{J}{2}\sum_{i=1}^{N}\left(\sigma_i^x\sigma_{i+1}^x+\sigma_i^y\sigma_{i+1}^y+\Delta\sigma_i^z\sigma_{i+1}^z\right),\nonumber\\
& \sigma^{x,y,z}_{N+1}=\sigma^{x,y,z}_1 \quad \text{and}\quad J>0\label{eq:XXZHamiltonian}
\end{align}
which is local and has only nearest neighbor interaction, i.e. each density Hamiltonian $ H_{i,i+1} $ acts on two consecutive sites. The Hamiltonian $ \mathbb{H} $ acts on $ N $ Hilbert spaces $ \cH\otimes...\otimes \cH $. Here each Hilbert space is a $ \mathbb{C}^2 $, therefore the matrix $  \mathbb{H} $ is $ 2^N\times 2^N $ and describes a spin chain where each site can have a spin up or a spin down. The name XXZ comes from the fact that the coefficient in front of $ \sigma_i^x\sigma_{i+1}^x $
and $ \sigma_i^y\sigma_{i+1}^y $ are equal. When the three coefficients are the same it is called  XXX while when there are three different coefficients it is called XYZ \footnote{We can think of the XXX as a spin chain without magnetic field, the XXZ with a magnetic field in the $ z $ direction, while XYZ has magnetic field in two transverse directions.}. 

Studying the eigenvalues and eigenvectors of $ \mathbb{H} $, one can see that depending on the value of $ \Delta $ this spin chain describes a ferromagnetic or antiferromagnetic model:

\textbf{a)} For $ \Delta>1 $, the system is ferromagnetic, i.e., our ground state is of the form\footnote{There are actually two possibilities, the state with all the spins down has same energy as the one with all spins up.} 

\begin{equation}
|0\rangle= |\uparrow\uparrow\uparrow...\uparrow\uparrow\rangle.
\end{equation} 

\textbf{b)} For $ \Delta<1 $ the system is antiferromagnetic and the ground state is of the form

\begin{equation}
|0\rangle=a_1 |\downarrow\uparrow\downarrow\uparrow...\downarrow\uparrow\rangle +a_2  |\downarrow\downarrow\uparrow\uparrow...\downarrow\uparrow\rangle+...
\end{equation}
i.e. for even $ N $, the ground state is a combination of all possible ways to distribute half of the spins down and half of the spins up in the chain. For odd $ N $ there are two possible ground states: one with $\frac{N-1}{2}$ spins up and $\frac{N+1}{2}$ spins down, and the other with the opposite configuration.

The claims a) and b) can be checked easily by computing the eigenvalues and eigenvectors of the Hamiltonian for a few number of sites. Let us plot the possible energies as a function of $ \Delta $ for a spin chain with $ N=3 $ to see that there is really something happening for $ \Delta=1 $

\begin{figure}[H]
\hspace{3cm}	\includegraphics[width=10cm]{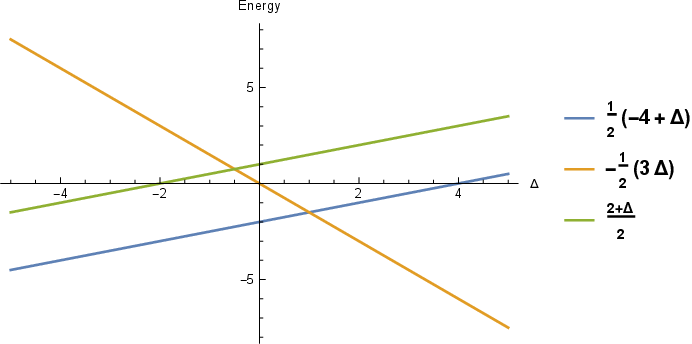}
\caption{Energies for the XXZ Hamiltonian for 3 sites.}
\end{figure}
\noindent where we can clearly see that for $ \Delta>1 $ the ground state energy is $E_0=-\frac{3\Delta}{2}  $ while for $ \Delta<1 $ the ground state energy is $ E_0=\frac{\Delta-4}{2} $. 
By computing the eigenvectors corresponding to these eigenvalues one can see that for $ \Delta>1  $ the ground state is ferromagnetic, and for $ \Delta<1 $ it is antiferromagnetic. We leave this calculation as an exercise.

\begin{tcolorbox}
	
	\paragraph{Exercise 14:} Compute the eigenvalues and eigenvectors of $ \mathbb{H} $ for two and three sites (in Mathematica) and convince yourself of the claims a) and b) above. What happens with the degeneracies for $ \Delta=1 $?
	
\end{tcolorbox}
This is a very interesting model which happens to be integrable. We can construct its Hamiltonian and all the other conserved charges in a systematic way using a quantized version of the R-matrix. In order to that we will use again the XXZ spin chain as an example.

\subsection{Quantum R-matrix and the quantum Yang-Baxter equation}\label{subsec:transfer}

As in classical integrability, a quantum integrable system is also characterized by a set of conserved quantities.  As expected, in quantum integrability we substitute the Poisson bracket by the commutator and have that the charges satisfy

\begin{equation}
\left[\mathbb{Q}_i,\mathbb{Q}_j\right]=0, \quad i,j=1,...
\label{eq:chargescommuting}
\end{equation}
We will in this section explain how to construct these charges starting from a quantum R-matrix.

The \textbf{ quantum $ \mathbf{R} $-matrix}  can be thought of as a mathematical object $ R(z_1,z_2) $ which maps $ \text{End}(\cH\otimes \cH)\mapsto \text{End}(\cH\otimes \cH) $, where $ z_i $ are the spectral parameters and each $ \cH $ is a Hilbert space. 

We will give later a nice interpretation for both $ R $ and $ z_i $. The $ R_{1,2}(z_1-z_2) $ can be represented as two lines crossing in the following way 

\begin{figure}[H]
	\hspace{7cm}	\includegraphics[width=2cm]{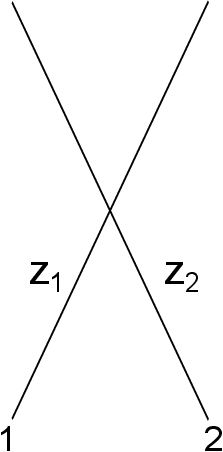}.
	\caption{Graphical representation of the R-matrix $ R(z_1,z_2) $}
\end{figure}

The R-matrix is defined as the solution of the quantum Yang-Baxter equation (qYBE) given by
\begin{equation}
R_{12}(z_1,z_2)R_{13}(z_1,z_3)R_{23}(z_2,z_3)=R_{23}(z_2,z_3)R_{13}(z_1,z_3)R_{12}(z_1,z_2)
\label{eq:qYBEnondif}
\end{equation}
where $ R_{ij} $ maps $ \text{End}(\cH\otimes \cH\otimes \cH) \mapsto \text{End}(\cH\otimes \cH\otimes \cH) $. The qYBE can be graphically represented by

 \begin{figure}[H]
 	\hspace{5cm}	\includegraphics[width=8cm]{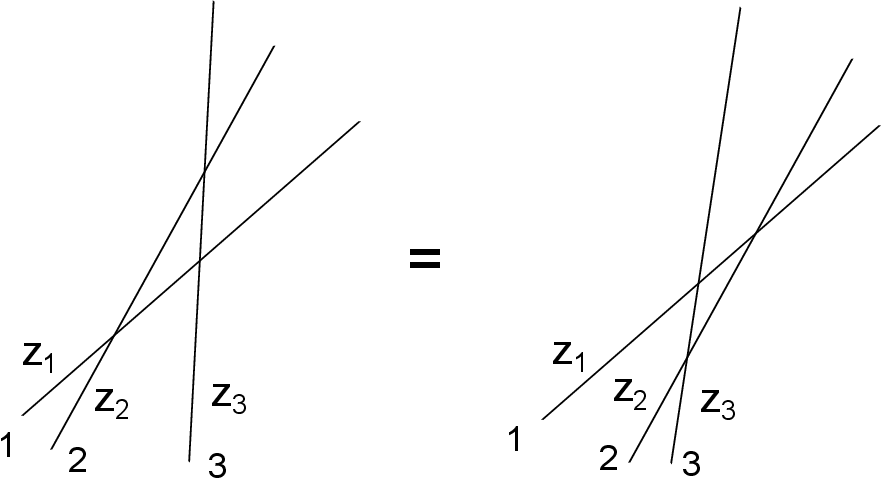}
 	\caption{Graphical representation of the qYBE equation. Each line has a space and a spectral parameter associated to it. And every time two lines cross we have an R-matrix}.
 \end{figure}

We will however focus here in the particular cases where
\begin{equation}
R_{i,j}(z_i,z_j)=R_{i,j}(z_i-z_j).
\label{eq:diffformR}
\end{equation}
R-matrices satisfying this property are called \textbf{difference form} R-matrices\footnote{R-matrices which do not satisfy this property are called \textbf{non-difference form} R-matrices. Examples are the Hubbard model and the AdS$ _5 \times $S$ ^5 $ R-matrix \cite{essler2005one,Beisert:2010jr}.}
	 
The qYBE considering \eqref{eq:diffformR} can be written as 
\begin{equation}
R_{12}(z_1-z_2)R_{13}(z_1)R_{23}(z_2)=R_{23}(z_2)R_{13}(z_1)R_{12}(z_1-z_2)
\label{eq:qYBE}
\end{equation}
where in addition to consider the condition \eqref{eq:diffformR} we also performed the transformations $ z_1\rightarrow z_1+z_3 $ and $ z_2\rightarrow z_2+z_3 $ making there fore the $ z_3 $ disappear. In this way qYBE depends only on two variables.

\subsection{Obtaining the classical r-matrix from the quantum R-matrix}\label{subsec:quantumRtoclassicalr}

The quantum R-matrix is related to its classical counterpart by the following expansion

\begin{equation}
R_{i,j}(z)=\kappa(z)\left(\mathbb{I}_{i,j}+\hbar \,r_{i,j}(z)+\mathcal{O}(\hbar^2)\right),
\label{eq:expansionRtor}
\end{equation}
where $ \kappa(z) $ is just a scalar function of $ z $.

By substituting this expansion in the quantum Yang-Baxter equation \eqref{eq:qYBE} and looking at the second order in $ \hbar $ we recover the classical Yang-Baxter equation \eqref{eq:cYBElambda}.
Let us now prove this claim by computing the lhs of qYBE

\begin{align}
\text{lhs}&=R_{12}(z_1-z_2)R_{13}(z_1)R_{23}(z_2)\nonumber\\
& =\kappa(z_1-z_2)\kappa(z_1)\kappa(z_2)\left(\mathbb{I}_{1,2}+\hbar \,r_{1,2}(z_1-z_2)+\mathcal{O}(\hbar^2)\right)\times\nonumber\\
&\hspace{0.5cm}\times \left(\mathbb{I}_{1,3}+\hbar \,r_{1,3}(z_1)+\mathcal{O}(\hbar^2)\right)\left(\mathbb{I}_{2,3}+\hbar \,r_{2,3}(z_2)+\mathcal{O}(\hbar^2)\right)\\
&=\kappa(z_1-z_2)\kappa(z_1)\kappa(z_2)\left[\hbar^0\,\mathbb{I}_{1,2}\mathbb{I}_{1,3}\mathbb{I}_{2,3}+\hbar^1 \left(r_{1,2}(z_1-z_2)+r_{1,3}(z_1)+r_{2,3}(z_2)\right)+\right.\nonumber\\
&\hspace{0.5cm}\left.+\hbar^2\left(r_{1,2}(z_1-z_2)r_{1,3}(z_1)+r_{1,2}(z_1-z_2)r_{2,3}(z_2)+r_{1,3}(z_1)r_{2,3}(z_2)\right)+\mathcal{O}(\hbar^3)\right]
\label{eq:lhsYBE}
\end{align}
and also its rhs
\begin{align}
\text{rhs}&=R_{23}(z_2)R_{13}(z_1)R_{12}(z_1-z_2)\nonumber\\
&=\kappa(z_1-z_2)\kappa(z_1)\kappa(z_2)\left(\mathbb{I}_{2,3}+\hbar \,r_{2,3}(z_2)+\mathcal{O}(\hbar^2)\right)\times\nonumber\\
&\hspace{0.5cm}\times\left(\mathbb{I}_{1,3}+\hbar \,r_{1,3}(z_1)+\mathcal{O}(\hbar^2)\right)\left(\mathbb{I}_{1,2}+\hbar \,r_{1,2}(z_1-z_2)+\mathcal{O}(\hbar^2)\right),\\
&=\kappa(z_1-z_2)\kappa(z_1)\kappa(z_2)\left[\hbar^0\,\mathbb{I}_{2,3}\mathbb{I}_{1,3}\mathbb{I}_{1,2}+\hbar^1 \left(r_{2,3}(z_2)+r_{1,3}(z_1)+r_{1,2}(z_1-z_2)\right)+\right.\nonumber\\
&\hspace{0.5cm}\left.+\hbar^2\left(r_{2,3}(z_2)r_{1,3}(z_1)+r_{2,3}(z_2)r_{1,2}(z_1-z_2)+r_{1,3}(z_1)r_{1,2}(z_1-z_2)\right)+\mathcal{O}(\hbar^3)\right]
\label{eq:rhsYBE}
\end{align}
and then we compare them order by order. In orders zero and one in $ \hbar $ both sides are trivially equal. Now, by requiring the order $ \hbar^2 $ match we obtain
\begin{align}
& r_{1,2}(z_1-z_2)r_{1,3}(z_1)+r_{1,2}(z_1-z_2)r_{2,3}(z_2)+r_{1,3}(z_1)r_{2,3}(z_2)=\nonumber\\
&\hspace{2cm}=r_{2,3}(z_2)r_{1,3}(z_1)+r_{2,3}(z_2)r_{1,2}(z_1-z_2)+r_{1,3}(z_1)r_{1,2}(z_1-z_2)\\
& \left[r_{1,2}(z_1-z_2),r_{1,3}(z_1)\right]+\left[r_{1,2}(z_1-z_2),r_{2,3}(z_2)\right]+\left[r_{1,3}(z_1),r_{2,3}(z_2)\right]=0
\end{align}
which corresponds exactly to the classical Yang-Baxter \eqref{eq:cYBElambda2}.

\begin{tcolorbox}
	\paragraph{Exercise 15:} Consider the quantum R-matrices for  $  A_1^{(1)} $ (given by expression \eqref{eq:XXZRmatrix}) and $ A_2^{(1)} $  given by
	\hspace{-0.1cm}\begin{equation}
	R(z)=\left(
	\begin{array}{ccccccccc}
	g(z ) & 0 & 0 & 0 & 0 & 0 & 0 & 0 & 0 \\
	0 & f(z ) & 0 & e^{-\frac{z}{2} }h(z ) & 0 & 0 & 0 & 0 & 0 \\
	0 & 0 & f(z ) & 0 & 0 & 0 & e^{-\frac{z}{2} }h(z ) & 0 & 0 \\
	0 & e^{\frac{z}{2} } h(z ) & 0 & f(z ) & 0 & 0 & 0 & 0 & 0 \\
	0 & 0 & 0 & 0 & g[z] & 0 & 0 & 0 & 0 \\
	0 & 0 & 0 & 0 & 0 & f(z ) & 0 & e^{-\frac{z}{2} }h(z ) & 0 \\
	0 & 0 & e^{\frac{z}{2}} h(z ) & 0 & 0 & 0 & f(z ) & 0 & 0 \\
	0 & 0 & 0 & 0 & 0 & e^{\frac{z}{2} } h(z ) & 0 & f(z ) & 0 \\
	0 & 0 & 0 & 0 & 0 & 0 & 0 & 0 & g(z ) \\
	\end{array}
	\right)
	\label{eq:A21quantum}
	\end{equation}
	where 
	\begin{align}
	&f(z)=2\,e^{\frac{z+4\eta}{2}}\sinh\left(\frac{z}{2}\right),\nonumber\\ 	&g(z)=2\,e^{\frac{z+4\eta}{2}}\sinh\left(\frac{z-4\eta}{2}\right),\nonumber\\ &h(z)=-2\,e^{\frac{z+4\eta}{2}}\sinh\left(2\eta\right)
	\end{align}
	and the anisotropy parameter $ \eta $ is related to $ \hbar $. Check that expanding around $ \eta=0 $, i.e.
	
	\begin{equation}
	R(z)= \alpha(z)\mathbb{I}+\beta(z)r(z)\eta+\mathcal{O}(\eta^2),
	\end{equation} 
	for some value of $ \alpha(z) $ and $ \beta(z) $, the $ r(z) $ correspond exactly to their classical version given in equations \eqref{eq:C11classical} and \eqref{eq:A21classical}.
	\footnote{In comparison with \cite{Jimbo:1985ua}) we have $ x\rightarrow e^z $ and $ k\rightarrow e^{2\,\eta} $. This is because Jimbo uses YBE in a form depending on $ R_{12}(z_1/z_2) $ instead of $ R_{12}(z_1-z_2) $.}
	
\end{tcolorbox}

\subsection{The XXZ R-matrix}\label{subsec:XXZRmatrix}

The solutions of the YBE with certain symmetries such as simple Lie algebras (XXX is an example) and affine Lie algebras (XXZ is an example) were found long ago \cite{Jimbo:1985ua,Kuniba:1991yd, Bazhanov:1984gu,Bazhanov:1986mu}. But any regular solution of the Yang-Baxter equation can be used to construct a closed spin chain with a local Hamiltonian through the method we explain now\footnote{For open spin chains we need one more ingredient, the reflection matrices $ K $ which describe the boundary conditions of the model \cite{Sklyanin:1988yz}. They satisfy together with the R-matrix the so called reflection equation or Boundary Yang-Baxter equation (BYBE).}. We will, however, focus our attention on the XXZ spin chain, whose R-matrix is given by

\begin{equation}
R(z)=\begin{pmatrix}
\sinh(z+\eta) & 0 & 0 & 0 \\
0 & \sinh z & \sinh \eta & 0\\
0 & \sinh\eta & \sinh z & 0\\
0 & 0 & 0 & \sinh(z+\eta) 
\end{pmatrix}
\label{eq:XXZRmatrix}.
\end{equation} 
For this model each Hilbert space $ \cH= \mathbb{C}^2 $.
Notice that in addition to the spectral parameter, we also have another parameter $ \eta $. Actually, all the R-matrices for affine Lie algebras are trigonometric and have this extra parameter $ \eta $. In the case of XXZ the symmetry is $ A_1^{(1)} $ (or $ \widehat{su}(2) $). If we carefully take the limit $  \eta\rightarrow 0 $ we obtain the XXX R-matrix instead. This corresponds at the level of the Hamiltonian \eqref{eq:XXZHamiltonian}, to sending $ \Delta\rightarrow 1 $. This classification is directly related to classification by Belavin and Drinfeld \cite{Belavin1982} discussed in the classical part of these notes.

\begin{tcolorbox}
	
	\paragraph{Exercise 16:} Check that the XXZ R-matrix satisfies qYBE.
	
\end{tcolorbox}

\begin{tcolorbox}
	\paragraph{Exercise 17:} Check that if you multiply the XXZ R-matrix by $ 1/\sinh(z+\eta) $, then do $ z\rightarrow z\, \epsilon $ and $ \eta\rightarrow i\, \epsilon  $ and then take the limit $ \epsilon\rightarrow 0 $ you obtain an R-matrix that is proportional to 
	
	\begin{equation}
	R(z)=\begin{pmatrix}
	z +i & 0 & 0 & 0 \\
	0 & z & i & 0\\
	0 & i & z & 0\\
	0 & 0 & 0 & z+i 
	\end{pmatrix}
	\end{equation}
	which is indeed the XXX R-matrix. 

\end{tcolorbox}

We can actually think R-matrix as the most important object in a quantum integrable system, since once we have it, we can systematically construct all the conserved charges.

The R-matrix \eqref{eq:XXZRmatrix} for XXZ model satisfies a regularity condition

\begin{equation}
R (0)=\sinh\eta P
\label{eq:regularity}
\end{equation}
which is fundamental to the construction of a local Hamiltonian; and unitarity
\begin{equation}
R_{\alpha\beta}(z)R_{\beta\alpha}(-z)=g(z)\mathbb{I}_{\alpha\beta},
\label{eq:unitarity}
\end{equation}
where 
\begin{equation}
g(z)=\sinh\left(\eta-z\right)\sinh\left(\eta+z\right).
\label{eq:gdefinition}
\end{equation}

\subsection{The Lax operator}\label{subsec:LaxQuantum}

As in the classical part, we will have a Lax operator. It can be interpreted as the transport between two consecutive sites

\begin{equation}
|v_{i+1}\rangle=\mathcal{L}_{i}|v_i\rangle. 
\label{eq:v2Lv1}
\end{equation}

Such object is defined by the RLL equation:

\begin{equation}
R_{\alpha\beta}(z_\alpha-z_\beta)\mathcal{L}_{\alpha,j}(z_\alpha)\mathcal{L}_{\beta,j}(z_\beta)=\mathcal{L}_{\beta,j}(z_\beta)\mathcal{L}_{\alpha,j}(z_\alpha)R_{\alpha\beta}(z_\alpha-z_\beta)
\label{eq:RLL}
\end{equation}
The $ \mathcal{L}_{\alpha,j}(z_\alpha) $ operator maps $ \text{End}(\mathcal{V}\otimes \mathcal{H})\mapsto \text{End}(\mathcal{V}\otimes \mathcal{H}) $, where a $ \cH $ ( $= \mathbb{C}^2 $ for XXZ) is a physical space  (indicated by $ j $) and $ \mathcal{V} $ (also equal to $ \mathbb{C}^2 $ for XXZ) is an auxiliary space  indicated by $ \alpha $.  The inclusion of this auxiliary space can seem a bit arbitrary at this point. We will soon however, that it allows us to construct the generating function of all conserved charges when we trace it out in the monodromy matrix. The Lax operator can be represented as 
\begin{figure}[H]
	\hspace{5.5cm}	\includegraphics[width=4cm]{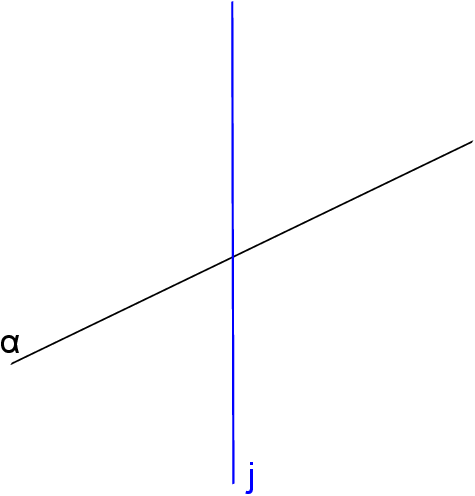}
	\caption{Graphical representation of the Lax operator. The blue line represents the physical space while the black line represents the auxiliary space.}
\end{figure}

\noindent where the blue and black lines represent physical and auxiliary space, respectively. With this, the RLL equation \eqref{eq:RLL} can be represented as

\begin{figure}[H]
	\hspace{4cm}	\includegraphics[width=7cm]{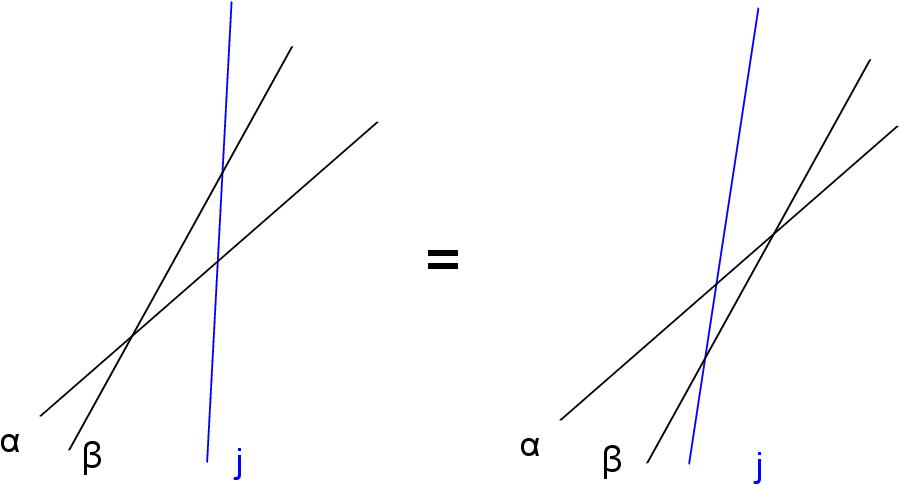}
	\caption{Graphical representation of the RLL relation \eqref{eq:RLL}}.\label{fig:RLL}
\end{figure}

 Given one R-matrix, solution of qYBE, usually there are many possible $ \cL_{\alpha,j} $ solutions for the  RLL  equation. One of its solutions is always the R-matrix itself, while the others correspond to different representations of the RLL algebra. In this lectures we are considering the case where $ \cL(z) $ is the same as $ R(z) $. 

So, assume $ \cL $ as the operator

\begin{align}
\mathcal{L}_{\alpha,j}(z)& =\frac{\sinh\eta}{2}\left(\sigma_\alpha^x\sigma_j^x+\sigma_\alpha^y\sigma_j^y\right)+\frac{\left(\sinh(z+\eta)-\sinh z\right)}{2}\sigma_\alpha^z\sigma_j^z+\nonumber\\
&\quad +\frac{\left(\sinh(z+\eta)+\sinh z\right)}{2}I_{\alpha,j},\label{eq:XXZLoperator1}\\
&=\begin{pmatrix}
\sinh\left(z+\frac{\eta}{2}\left(1+\sigma_j^z\right)\right) & \sigma_j^-\sinh\left(\frac{\eta}{2}\right)\\
\sigma_j^+\sinh\left(\frac{\eta}{2}\right)&\sinh\left(z+\frac{\eta}{2}\left(1-\sigma_j^z\right)\right)
\end{pmatrix} 
\label{eq:XXZLoperator2}
\end{align}
where $ \sigma^\pm=\frac{1}{2}\left(\sigma^x\pm i \sigma^y\right) $. 

Notice that the two ways of writing the $ \cL $ operator above are completely equivalent. Depending on what we are doing, one version or the other will be more convenient. As already mentioned, it also coincides with the R-matrix \eqref{eq:XXZRmatrix}.

\begin{tcolorbox}
	
	\paragraph{Exercise 18:} Convince yourself that $ \cL_{\alpha,j}(z) $ written as in \eqref{eq:XXZLoperator1} and in \eqref{eq:XXZLoperator2} are equivalent, as well as in \eqref{eq:XXZRmatrix}, and check that it satisfies the  RLL equation \eqref{eq:RLL}.
	
\end{tcolorbox}

\begin{tcolorbox}
	
	\paragraph{Exercise 19:} Prove that 
	\begin{align}
	& \cL_{\alpha,j}(0)\propto P_{\alpha,j},\label{eq:Lprop1}\\
	& \tr_{\alpha}\cL_{\alpha,j}(0)\propto I_j,\label{eq:Lprop2}.
	\end{align}
	These properties will play an important role in the construction of the Hamiltonian. 
	
\end{tcolorbox}

\subsection{The monodromy matrix}\label{subsec:monodromy}
Since $ \cL $  is responsible for transport among consecutive sites, we can use it also to transport among sites very far apart. We can define the monodromy matrix, from site $ 1 $ to site $ N $ as

\begin{equation}
\cT_\alpha(z)= \mathcal{L}_{\alpha,N}(z)\mathcal{L}_{\alpha,N-1}(z)...\mathcal{L}_{\alpha,2}(z)\mathcal{L}_{\alpha,1}(z)
\label{eq:monodromy}
\end{equation}
which can be represented as

\begin{figure}[H]
	\hspace{4cm}	\includegraphics[width=7cm]{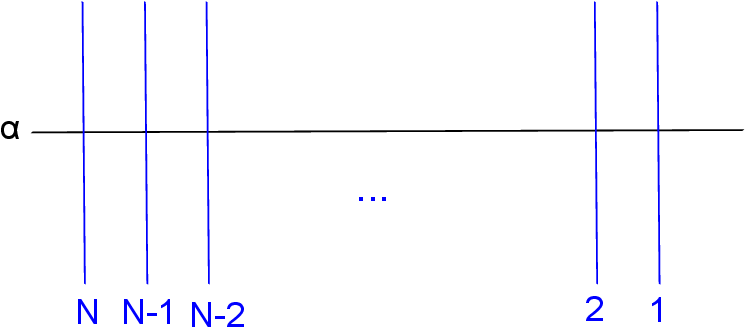}
	\caption{Graphical representation of the monodromy matrix}
\end{figure}

The monodromy matrix satisfies the following equation

\begin{equation}
R_{\alpha\beta}(z_\alpha-z_\beta)\cT_\alpha(z_\alpha)\cT_\beta(z_\beta)=\cT_\beta(z_\beta)\cT_\alpha(z_\alpha)R_{\alpha\beta}(z_\alpha-z_\beta).
\label{eq:RTT}
\end{equation}

Let us write the proof for two sites, but it can be easily generalized for any number of sites. The monodromy matrix is given by
\begin{equation}
\cT_\alpha(z_\alpha)=\cL_{\alpha,2}(z_\alpha)\cL_{\alpha,1}(z_\alpha).\\
\end{equation}
Substituting them in the equation \eqref{eq:RTT} we find
\begin{align}
R_{\alpha\beta}(z_\alpha-z_\beta)\cT_\alpha(z_\alpha)\cT_\beta(z_\beta)& =R_{\alpha\beta}(z_\alpha-z_\beta)\cL_{\alpha,2}(z_\alpha)\cL_{\alpha,1}(z_\alpha)\cL_{\beta,2}(z_\beta)\cL_{\beta,1}(z_\beta)\nonumber\\
& =R_{\alpha\beta}(z_\alpha-z_\beta)\cL_{\alpha,2}(z_\alpha)\cL_{\beta,2}(z_\beta)\cL_{\alpha,1}(z_\alpha)\cL_{\beta,1}(z_\beta)\nonumber\\
&= \cL_{\beta,2}(z_\beta)\cL_{\alpha,2}(z_\alpha)R_{\alpha\beta}(z_\alpha-z_\beta)\cL_{\alpha,1}(z_\alpha)\cL_{\beta,1}(z_\beta)\nonumber\\
& = \cL_{\beta,2}(z_\beta)\cL_{\alpha,2}(z_\alpha)\cL_{\beta,1}(z_\beta)\cL_{\alpha,1}(z_\alpha)R_{\alpha\beta}(z_\alpha-z_\beta)\nonumber\\
& = \cL_{\beta,2}(z_\beta)\cL_{\beta,1}(z_\beta)\cL_{\alpha,2}(z_\alpha)\cL_{\alpha,1}(z_\alpha)R_{\alpha\beta}(z_\alpha-z_\beta)\nonumber\\
&=\cT_\beta(z_\beta)\cT_\alpha(z_\alpha)R_{\alpha\beta}(z_\alpha-z_\beta)
\end{align}
where in the first and fourth steps we use the fact that operators acting on different sites commute (which you proved on section \ref{subsec:Kroneckerproduct}), steps two and three result from using the RLL equation \eqref{eq:RLL} and finally, the last step is just to recognize again the products as the monodromy matrices on sites $ \alpha $ and $ \beta $.

We can also prove the RTT relation \eqref{eq:RTT} for any number of sites using pictures. Given that $ \cT_\alpha $ and $ \cT_\beta $ act on the same physical spaces, i.e., both $ \cL_{\alpha,N} $ and $ \cL_{\beta,N} $ act on site $ N $; both $ \cL_{\alpha,N-1} $ and $ \cL_{\beta,N-1} $ act on site $ N-1 $, etc, we can draw the lhs for equation \eqref{eq:RTT} like

\begin{figure}[H]
	\hspace{5cm}	\includegraphics[width=7cm]{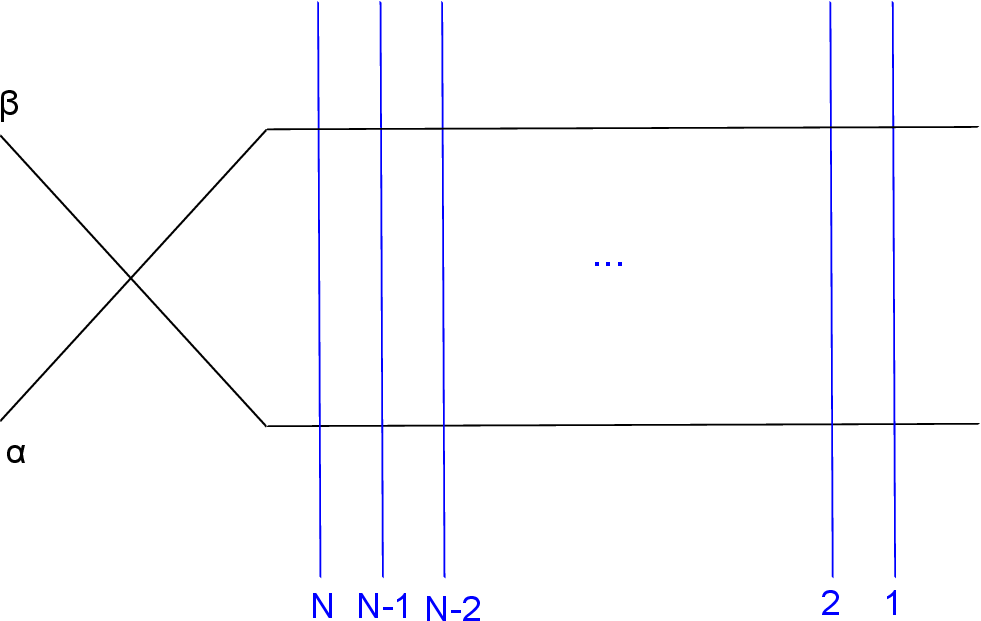}
	\caption{LHS of the RTT equation \eqref{eq:RTT}}
\end{figure}

\noindent and then using the figure \ref{fig:RLL} we can pass the $ \cL $ operators, two by two throught $ R $ until we obtain

\begin{figure}[H]
	\hspace{5cm}	\includegraphics[width=7cm]{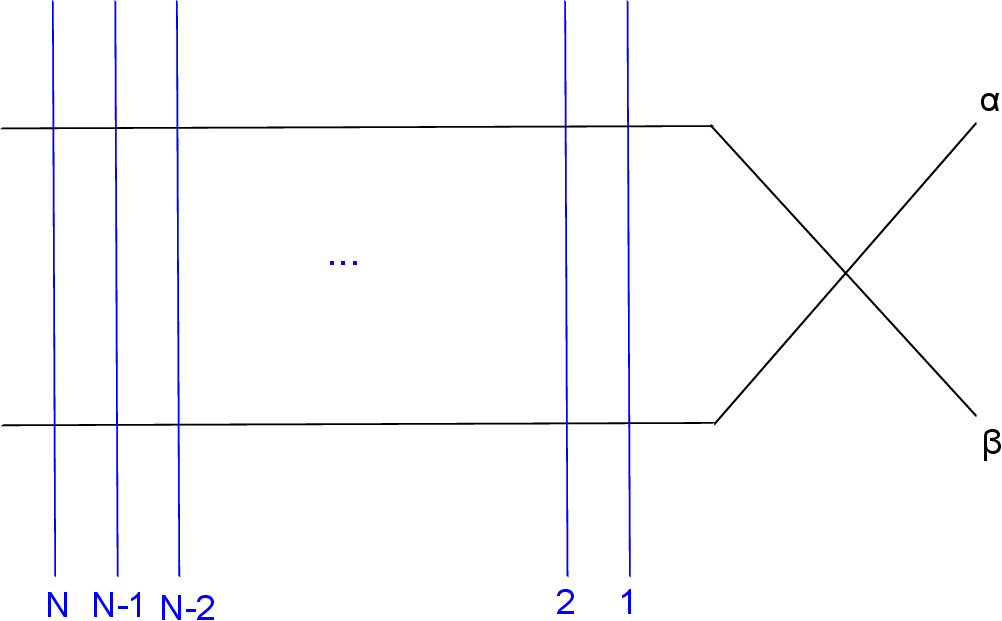}
	\caption{RHS of the RTT equation \eqref{eq:RTT}}
\end{figure}

\noindent which corresponds to rhs of equation \eqref{eq:RTT}.

\subsection{The transfer matrix}\label{subsec:Qtransfer}

Here the reason why the introduction of the auxiliary space makes sense becomes clearer. It happens that by tracing out such space we get a generating function of all the conserved charges: the\textbf{ transfer matrix} as we prove below.

The transfer matrix is then given by
\begin{equation}
t(z)=\text{tr}_{\alpha}\cT_\alpha(z)
\label{eq:transfer}
\end{equation}
which satisfies

\begin{equation}
\left[t(z_1),t(z_2)\right]=0.
\label{eq:commutativityproperty}
\end{equation}
The proof is actually very simple:

\begin{align}
t(z_\alpha)t(z_\beta)&=\text{tr}_\alpha\cT_\alpha(z_\alpha)\text{tr}_\beta\cT_\beta(z_\beta)\nonumber\\
& =\text{tr}_{\alpha\beta}\left(\cT_\alpha(z_\alpha)\cT_\beta(z_\beta)\right),\\
&=\frac{1}{g(z_\alpha-z_\beta)} \, \text{tr}_{\alpha\beta}\left(R_{\beta,\alpha}\left(z_\beta-z_\alpha\right)R_{\alpha,\beta}\left(z_\alpha-z_\beta\right)\cT_\alpha(z_\alpha)\cT_\beta(z_\beta)\right)\nonumber\\
& =\frac{1}{g(z_\alpha-z_\beta)}\,\text{tr}_{\alpha\beta}\left(R_{\beta,\alpha}\left(z_\beta-z_\alpha\right)\cT_\beta(z_\beta)\cT_\alpha(z_\alpha)R_{\alpha,\beta}\left(z_\alpha-z_\beta\right)\right)\nonumber\\
& =\frac{1}{g(z_\alpha-z_\beta)}\,\text{tr}_{\alpha\beta}\left(\cT_\alpha(z_\alpha)\cT_\beta(z_\beta)R_{\beta,\alpha}\left(z_\beta-z_\alpha\right)R_{\alpha,\beta}\left(z_\alpha-z_\beta\right)\right)\nonumber\\
&=\text{tr}_{\alpha\beta}\left(\cT_\beta(z_\beta)\cT_\alpha(z_\alpha)\right)\nonumber\\
&=\text{tr}_\beta\cT_\beta(z_\beta)\text{tr}_\alpha\cT_\alpha(z_\alpha)\nonumber\\
&=t(z_\beta)t(z_\alpha),
\end{align}
where $ g(z_\alpha-z_\beta) $ was defined in \eqref{eq:gdefinition}. The proof uses unitarity \eqref{eq:unitarity} and the RTT relation \eqref{eq:RTT}.

The transfer matrix can be represented as

\begin{figure}[H]
	\hspace{5cm}	\includegraphics[width=7cm]{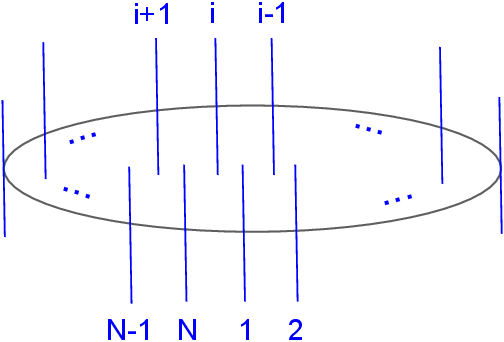}
	\caption{Graphical representation of the transfer matrix}
\end{figure}

Looking at the explicit form of $ \cL_{\alpha,j} $ it is easy to see that the transfer matrix can be written as a series in $ z $

\begin{equation}
\text{log}\,t(z)=\sum_{l}\mathbb{Q}_lz^l,
\end{equation}
which means that the charges are given by

\begin{equation}
\mathbb{Q}_{l+1}=\frac{d^{l}}{dz^{l}}\text{log}\,t(z)\Big|_{z=0},\quad l=0,1,...\,.
\end{equation}

Given, the commutativity property \eqref{eq:commutativityproperty}, it follows that

\begin{equation}
\left[\mathbb{Q}_l,\mathbb{Q}_m\right]=0, \quad \forall \quad l,m\in\mathbb{Z}^+.
\end{equation}

Let us now look a bit more at these charges. Let us start with $ \mathbb{Q}_1 $:

\begin{align}
\mathbb{Q}_1&=\text{log}\, t(0)=(\sinh\eta)^N P_{1,2}P_{2,3}...P_{N-1,N},\nonumber\\
&=i \hat{\mathbb{P}},
\end{align}
which corresponds to momentum. The second charge is

\begin{align}
\mathbb{Q}_2& =\frac{d}{dz}\text{log}\,t(z)\Big|_{z=0}\nonumber\\
&=t^{-1}(0)t'(0)
\label{eq:Q2XXZ}
\end{align}
where $ t'(0) $ means that we take the derivative with respect to the spectral parameter $ z $ and then do $ z\rightarrow 0 $.

In order to have something concrete, let us compute this for $ N=3 $. 
The transfer matrix for 3 sites is 

\begin{equation}
t(z)=\text{tr}_\alpha\left(\cL_{\alpha,3}(z)\cL_{\alpha,2}(z)\cL_{\alpha,1}(z)\right),
\end{equation}
so
\begin{align}
t(0)&=\text{tr}_\alpha\left(\cL_{\alpha,3}(0)\cL_{\alpha,2}(0)\cL_{\alpha,1}(0)\right),\nonumber\\
&=(\sinh\eta)^3\, \text{tr}_{\alpha}(P_{\alpha,3}P_{\alpha,2}P_{\alpha,1}),\nonumber\\
&=(\sinh\eta)^3 \left(\text{tr}_{\alpha}P_{\alpha_1}\right)P_{13}P_{12},\nonumber\\
&=(\sinh\eta)^3 P_{12}P_{23}
\end{align} 
where we use the property \eqref{eq:Lprop1}, then a few times \eqref{eq:PAPproperty} with $ B=P $, and finally we use $ \text{tr}_{\alpha}P_{\alpha,1}=I_1 $ (proved in an exercise in section \eqref{subsec:Kroneckerproduct}). The idea when computing the charges is always the same, to use permutation $ P_{\alpha,j} $ through property \eqref{eq:PAPproperty} to remove as many terms as possible from inside the trace. Let us now do this for $ t'(0) $ in \eqref{eq:Q2XXZ}

\begin{align}
\mathbb{Q}_2=t^{-1}(0)t'(0)&=\left(\sinh^{-3}\eta P_{23}P_{12}\right) \left(\tr_{\alpha}\cL_{\alpha,3}'(0)\cL_{\alpha,2}(0)\cL_{\alpha,1}(0)+\right.\nonumber\\
&\left.\quad +\tr_{\alpha}\cL_{\alpha,3}(0)\cL_{\alpha,2}'(0)\cL_{\alpha,1}(0) +\tr_{\alpha}\cL_{\alpha,3}(0)\cL_{\alpha,2}(0)\cL_{\alpha,1}'(0)\right)\label{eq:sum1}\\
&=\frac{1}{\sinh\eta}\left(P_{23}\cL_{23}'(0)+P_{12}\cL_{12}'(0)+P_{31}\cL_{31}'(0)\right)\label{eq:sum2}\\
&=\sum_{i=1}^{3}P_{i,i+1}\cL_{i,i+1}'(0)=\mathbb{H}.\label{eq:sum3}
\end{align}

We compute explicitly the first term in \eqref{eq:sum1} and leave the others as an exercise 

\begin{align}
& \left(\sinh^{-3}\eta P_{23}P_{12}\right) \left(\tr_{\alpha}\cL_{\alpha,3}'(0)\cL_{\alpha,2}(0)\cL_{\alpha,1}(0)\right)=\nonumber\\
& \hspace{2cm} = \left(\sinh^{-1}\eta P_{23}P_{12}\right) \left(\tr_{\alpha}\cL_{\alpha,3}'(0)P_{\alpha,2}(0)P_{\alpha,1}(0)\right)\\
& \hspace{2cm} =\left(\sinh^{-1}\eta P_{23}P_{12}\right)(\tr_{\alpha}P_{\alpha,1})\cL_{1,3}'(0)P_{12}\\
& \hspace{2cm} =\frac{1}{\sinh\eta}P_{23}P_{12}\cL_{1,3}'(0)P_{12}\\
& \hspace{2cm} =\frac{1}{\sinh\eta}P_{23}\cL_{23}'(0).
\end{align}

\begin{tcolorbox}
	
	\paragraph{Exercise 20:} Construct the two remaining terms and check that they are indeed equal to what is written in \eqref{eq:sum2}. 
	
\end{tcolorbox}

By doing a few cases, it is very easy to see that it generalizes to 

\begin{align}
\mathbb{H}&=\sum_{i=1}^{N}H_{i,i+1}=\sum_{i=1}^{N}\left(a\,P_{i,i+1}\cL_{i,i+1}'(0)+b\, I_{i,i+1}\right), \quad \text{with} \quad  H_{N,N+1}=H_{N,1},\label{eq:fullHline1}\\
&=-\frac{J}{2}\sum_{i=1}^{N}\left(\sigma_i^x\sigma_{i+1}^x+\sigma_i^y\sigma_{i+1}^y+\Delta\sigma_i^z\sigma_{i+1}^z\right).\label{eq:fullHline2}
\end{align}
We added the constants $ a $ and $ b $ in order to make \eqref{eq:fullHline1} match with \eqref{eq:XXZHamiltonian}. But, notice that we can always include such constants. The $ a $ is due to the fact that RLL is invariant under normalization of the Lax operator, while $ b $ just changes where the ``zero" of the energy is.

In order to go from equation \eqref{eq:fullHline1} to \eqref{eq:fullHline2} we computed explicitly $ \cL_{i,i+1}'(0) $, i.e. we took $ \cL_{\alpha,j}(z) $ from equation \eqref{eq:XXZLoperator1} with $ \alpha=i $ and $ j=i+1 $, applied the derivative with respect to $ z $, and at the end sent $ z $ to zero.

\begin{tcolorbox}
	\paragraph{Exercise 21:} By comparing \eqref{eq:fullHline1} with \eqref{eq:fullHline2} compute the value of $ a $, $ b $ and $ \Delta $ in terms of $ \eta $.
\end{tcolorbox}

Using a similar strategy one can continue to construct as many charges as wanted. In principle we could want to diagonalize many of them. 

The good news is that we do not have to! We can instead diagonalize the transfer matrix 

\begin{equation}
t(z)|\Lambda\rangle= \Lambda(z)|\Lambda\rangle
\end{equation}
and then the eigenvalues for all the charges can be computed by

\begin{equation}
\Lambda^{(Q_{n+1})}=\frac{d^{n}}{d z^{n}}\,\text{log}\Lambda(z)\Big|_{z=0}.
\end{equation}

One could reach this point and ask why we did not diagonalize the Hamiltonian directly, instead of constructing all this machinery. There are two reasons, one is that if a model is integrable it has many conserved charges and it is interesting to be able to construct them in a systematic way. The second reason is that although in theory one could indeed diagonalize the Hamiltonian directly, it is a very big matrix, whose size increases exponentially with the number of sites. Below, we will see that this machinery allows for the construction of a different method to diagonalize the Hamiltonian, the so called \textbf{Algebraic Bethe ansatz (ABA)}. Using this method the Hamiltonians for much bigger spin chains can be diagonalized.

\subsection{Algebraic Bethe ansatz (ABA)}\label{subsec:ABA}

There are several types of Bethe ansatz (Coordinate, algebraic, analytical, thermodynamic, nested algebraic, etc) and they all have advantages and disadvantages. Since we have been discussing the construction with Lax and transfer matrices, it is very natural that we decide here to present the algebraic Bethe ansatz. 

It is a very powerful tool that allows us to find both the eigenvalues and eigenvectors of the transfer matrix. But more than that, to find not only the spectrum of the Hamiltonian but the spectrum of all the conserved charges of our model in a simple way.

Let us now rewrite the monodromy matrix as

\begin{equation}
\mathcal{T}_\alpha(z)=\mathcal{L}_{\alpha,N}(z)\mathcal{L}_{\alpha,N-1}(z)\,...\,\mathcal{L}_{\alpha,1}(z)=\begin{pmatrix}
\mathcal{A}(z) & \mathcal{B}(z)\\
\mathcal{C}(z) & \mathcal{D}(z)
\end{pmatrix}
\label{eq:TABCD}
\end{equation}
where the elements $ \mathcal{A}(z),\,\mathcal{B}(z),\,\mathcal{C}(z),\,\mathcal{D}(z) $ are  $ 2^N\times 2^N $ matrices acting on the physical Hilbert spaces $ \mathbb{C}^2$. In this way, the transfer matrix is nothing more than

\begin{equation}
t(z)=\text{tr}_\alpha \mathcal{T}_\alpha(z)=\mathcal{A}(z)+\mathcal{D}(z).
\label{eq:t=A+D}
\end{equation}

Consider the ferromagnetic pseudo-vacuum 

\begin{equation}
|0\rangle =\begin{pmatrix}
1\\
0
\end{pmatrix}\otimes\begin{pmatrix}
1\\
0
\end{pmatrix}\otimes \begin{pmatrix}
1\\
0
\end{pmatrix}\otimes \,...\,\otimes \begin{pmatrix}
1\\
0
\end{pmatrix}.
\label{eq:pseudovacuum}
\end{equation}

For XXZ, by computing the monodromy matrix $ \cT_a(z) $ for a few sites, it is easy to find that

\begin{align}
& \mathcal{A}(z)\,|0\rangle=\sinh^N\left(z+\eta\right)\,|0\rangle\label{eq:Aaction},\\
& \mathcal{D}(z)\,|0\rangle=\sinh^N\left(z\right)\,|0\rangle,\label{eq:Daction}\\
& \mathcal{C}(z)\,|0\rangle=0\label{eq:Caction},\\
& \mathcal{B}(z)\,|0\rangle=\sinh\eta\,\sum_{i=1}^{N}\sinh^{N-i}\left(z+\eta\right)\sinh^{i-1}z\,|i\rangle\label{eq:Baction}
\end{align}
where 

\begin{equation}
|i\rangle=\begin{pmatrix}
1\\
0
\end{pmatrix}\otimes\begin{pmatrix}
1\\
0
\end{pmatrix}\otimes \,...\,\otimes\underbrace{\begin{pmatrix}
0 \\ 1
\end{pmatrix}}_{i-\text{th site}} \otimes \,...\,\otimes \begin{pmatrix}
1\\
0
\end{pmatrix}
\end{equation}
i.e. we have spins up in all the sites except in the site $ i $ where there is a spin down.

\begin{tcolorbox}
	\paragraph{Exercise 22:} Compute $ \mathcal{A}(z) $, $ \mathcal{B}(z) $, $ \mathcal{C}(z) $ and $ \mathcal{D}(z) $ for a few sites and then check that equations \eqref{eq:Aaction}-\eqref{eq:Baction} are correct. 
\end{tcolorbox}

Notice that we start with the lowest energy state and by applying the $ \mathcal{B}(z) $ operator to it we create an excited state. So, the $ \mathcal{B}(z) $ works as some creation operator.

The existence of a pseudo-vacuum \eqref{eq:pseudovacuum} is fundamental in this construction since everything is built starting from it.

We define excited \textbf{Bethe states} as  

\begin{equation}
|\Lambda(z_1,...,z_m)\rangle =\mathcal{B}(z_1)\mathcal{B}(z_2)...\mathcal{B}(z_m)|0\rangle. 
\label{eq:BB...B}
\end{equation}
For general values of $ \{z_i\} $ the state $  |\Lambda(z_1,...,z_m)\rangle $ is not an eigenvector of the transfer matrix. Let us see now how to discover for which values of $ \{z_i\} $  this state is an eigenstate of $ t(z) $.

We know that the transfer matrix is given by \eqref{eq:t=A+D} and we know how $ \mathcal{A}(z) $ and $ \mathcal{D}(z) $ act on the pseudo-vacuum $ |0\rangle $. 
Therefore when computing 
\begin{align}
t(u)|\Lambda(z_1,...,z_m)\rangle&=t(u)\mathcal{B}(z_1)\mathcal{B}(z_2)...\mathcal{B}(z_m)|0\rangle,\nonumber\\
&=(\mathcal{A}(z)+\mathcal{D}(z))\mathcal{B}(z_1)\mathcal{B}(z_2)...\mathcal{B}(z_m)|0\rangle
\end{align}
the strategy is to manage to pass $ \mathcal{A} (z)$ and $ \mathcal{D}(z) $ through all those $\mathcal{B}(z_i) $, so they can act directly on $ |0\rangle $. It happens that the formalism itself has the perfect tool for this operation: equation \eqref{eq:RTT}. By substituting $ \cT(z) $ in the form \eqref{eq:TABCD} in the equation \eqref{eq:RTT} one obtains several commutation relations. The relevant ones for us are the ones involving $ \mathcal{A} $ and $ \mathcal{B} $, $ \mathcal{D} $ and $ \mathcal{B} $ and $ \mathcal{B} $ and $ \mathcal{B} $ which are 
\begin{align}
& \left[\mathcal{B}(z),\mathcal{B}(z_1)\right]=0,\label{eq:commutationBB}\\
& \mathcal{A}(z)\mathcal{B}(z_1)=\frac{\sinh\eta}{\sinh\left(z-z_1\right)}\mathcal{B}(z)\mathcal{A}(z_1)+\frac{\sinh\left(z-z_1-\eta\right)}{\sinh\left(z-z_1\right)}\mathcal{B}(z_1)\mathcal{A}(z),\label{eq:commutationAB}\\
& \mathcal{D}(z)\mathcal{B}(z_1)=-\frac{\sinh\eta}{\sinh\left(z-z_1\right)}\mathcal{B}(z)\mathcal{D}(z_1)+\frac{\sinh\left(z-z_1+\eta\right)}{\sinh\left(z-z_1\right)}\mathcal{B}(z_1)\mathcal{D}(z)\label{eq:commutationDB}
\end{align}

\begin{tcolorbox}
	
	\paragraph{Exercise: 23} Obtain the commutation relations \eqref{eq:commutationBB}-\eqref{eq:commutationDB}.
	
\end{tcolorbox}

Ideally, what we want is 

\begin{equation}
t(z)\mathcal{B}(z_1)\mathcal{B}(z_2)...\mathcal{B}(z_m)|0\rangle=\Lambda(z,z_1,...,z_m)\mathcal{B}(z_1)\mathcal{B}(z_2)...\mathcal{B}(z_m)|0\rangle
\label{eq:eigenvalueeqfort}
\end{equation}
because this means that $ \Lambda(z,z_1,...,z_m) $ is the eigenvalue of $ t(z) $.

So, the first step is to pass $ \cA (z)$ and $ \cD (z)$ through $ \cB(z_1) $ using \eqref{eq:commutationAB} and $ \eqref{eq:commutationDB} $
\begin{align}
t(u)|\Lambda(z_1,...,z_m)\rangle&=(\mathcal{A}(z)+\mathcal{D}(z))\mathcal{B}(z_1)\mathcal{B}(z_2)...\mathcal{B}(z_m)|0\rangle\nonumber\\
& = \frac{\sinh\eta}{\sinh(z-z_1)}\cB(z)\cA(z_1)\cB(z_2)...\cB(z_m)|0\rangle+\nonumber\\
&\quad-\frac{\sinh\eta}{\sinh(z-z_1)}\cB(z)\cD(z_1)\cB(z_2)...\cB(z_m)|0\rangle+\nonumber\\&\quad+ \frac{\sinh(z-z_1-\eta)}{\sinh(z-z_1)}\cB(z_1)\cA(z)\cB(z_2)...\cB(z_m)|0\rangle+\nonumber\\&\quad+ \frac{\sinh(z-z_1+\eta)}{\sinh(z-z_1)}\cB(z_1)\cD(z)\cB(z_2)...\cB(z_m)|0\rangle
\end{align}

After this first step what we get are two types of terms

\paragraph{1)} the first and second terms are what we call ``unwanted terms". Notice that they have a $ \cB(z) $ instead of $ \cB(z_1) $, so in principle we can continue to pass $ \cA(z_1) $ and $ \cD(z_1) $ through the $ \cB(z_i) $ but it is impossible to rewrite these terms in the form \eqref{eq:BB...B}, therefore this type of term has to be canceled in order for us to obtain \eqref{eq:eigenvalueeqfort}. Let us keep this term for now

\begin{align}``\text{unwanted term  1}"\,&=\frac{\sinh\eta}{\sinh(z-z_1)}\cB(z)\cA(z_1)\cB(z_2)...\cB(z_m)|0\rangle +\nonumber\\
&\quad-\frac{\sinh\eta}{\sinh(z-z_1)}\cB(z)\cD(z_1)\cB(z_2)...\cB(z_m)|0\rangle
\end{align}
 and deal with it later.

\paragraph{2)} the third and fourth terms are called ``wanted terms" and have $ \cB(z_1) $ so, we continue with these ones that can, potentially, give us the eigenvalue:

\begin{align}
``\text{wanted term}"\,&=\frac{\sinh(z-z_1-\eta)}{\sinh(z-z_1)}\cB(z_1)\cA(z)\cB(z_2)...\cB(z_m)|0\rangle+\nonumber\\
&\quad +\frac{\sinh(z-z_1+\eta)}{\sinh(z-z_1)}\cB(z_1)\cD(z)\cB(z_2)...\cB(z_m)|0\rangle
\end{align}

Applying again the commutation relations \eqref{eq:commutationAB}-\eqref{eq:commutationDB} we obtain 

\begin{align}
``\text{wanted term}"\,&=\frac{\sinh(z-z_1-\eta)}{\sinh(z-z_1)}\frac{\sinh(z-z_2-\eta)}{\sinh(z-z_2)}\cB(z_1)\cB(z_2)\cA(z)\cB(z_3)...\cB(z_m)|0\rangle+\nonumber\\
&\quad +\frac{\sinh(z-z_1+\eta)}{\sinh(z-z_1)}\frac{\sinh(z-z_2+\eta)}{\sinh(z-z_2)}\cB(z_1)\cB(z_2)\cD(z)\cB(z_3)...\cB(z_m)|0\rangle+\nonumber\\
&\quad+ \text{``unwanted terms 2"}.
\end{align}

By continuing repeatedly doing this we obtain

\begin{align}
t(z)\mathcal{B}(z_1)\mathcal{B}(z_2)...\mathcal{B}(z_m)|0\rangle&=\Lambda(z,\{z_1,...,z_m\})\mathcal{B}(z_1)\mathcal{B}(z_2)...\mathcal{B}(z_m)|0\rangle+\nonumber\\
&+\sum_{i=1}^{m}\left(\mathcal{M}^{(\cA)}_i(z,\left\{z_1,z_2,...\right\})+\mathcal{M}^{(\cD)}_i(z,\left\{z_1,z_2,...\right\})\right)\times\nonumber\\
& \quad \times \mathcal{B}(z)\mathcal{B}(z_1)\mathcal{B}(z_2)...\hat{\mathcal{B}}(z_i)...\mathcal{B}(z_m)|0\rangle
\end{align}
where 
\begin{equation}
\Lambda(z,\{z_1,...,z_m\})=\prod_{i=1}^{m}\left(\sinh^N\left(z+\eta\right)\frac{\sinh(z-z_i-\eta)}{\sinh(z-z_i)}+\sinh^N\left(z\right)\frac{\sinh(z-z_i+\eta)}{\sinh(z-z_i)}\right)
\label{eq:eigenvalue}
\end{equation}
while the sum is composed by the ``unwanted terms" coming from the commutations involving $ \cA $ represented by $\mathcal{M}_i^{(A)}$ and the ones coming from $ \cD $ represented by $ \mathcal{M}_i^{(D)} $. Since we want to find the eigenvalues of the transfer matrix, we need that the ``unwanted terms" cancel, so

\begin{equation}
\mathcal{M}_i^{(D)}=-\mathcal{M}_i^{(A)} \quad \text{for} \quad i=1,...,m.
\label{eq:MD=-MA}
\end{equation}

In principle computing $ \mathcal{M}_i $ sounds hard, but there is a trick. Notice that $ \mathcal{M}_1^{(A)} $ and $ \mathcal{M}_1^{(D)} $ are much simpler than the rest, because they only come from the ``unwanted term 1". So by continuing to apply the commutation relations, keeping only the terms that have $ \cA(z_1) $ and $ \cD(z_1) $ we obtain 

\begin{align}
&\mathcal{M}_1^{(A)}=\sinh^N\left(z_1+\eta\right)\frac{\sinh\,\eta}{\sinh\left(z-z_1\right)}\prod_{k=2}^{m}\frac{\sinh(z_1-z_k-\eta)}{\sinh(z_1-z_k)},\\
&\mathcal{M}_1^{(D)}=-\sinh^N\left(z_1\right)\frac{\sinh\,\eta}{\sinh\left(z-z_1\right)}\prod_{k=2}^{m}\frac{\sinh(z_1-z_k+\eta)}{\sinh(z_1-z_k)}.
\end{align}
Well, now comes a very nice part. Because the $ \cB(z_i) $ commute (as we found on \eqref{eq:commutationBB}) we could have started with $ \cB(z_2) $ first, or  $ \cB(z_3) $, etc and computed the corresponding $ \mathcal{M}_2^{(A)} $ and  $ \mathcal{M}_2^{(D)} $ or $ \mathcal{M}_3^{(A)} $ and $ \mathcal{M}_3^{(D)} $ using this. This means that all the $ \mathcal{M}_i $ are basically like $ \mathcal{M}_1 $, but changing all the indices 1 by $ i $. So, 

\begin{align}
&\mathcal{M}_i^{(A)}=\sinh^N\left(z_i+\eta\right)\frac{\sinh\,\eta}{\sinh\left(z-z_i\right)}\prod_{k\neq i}^{m}\frac{\sinh(z_i-z_k-\eta)}{\sinh(z_i-z_k)},\label{eq:BEoffshellA}\\
&\mathcal{M}_i^{(D)}=-\sinh^N\left(z_i\right)\frac{\sinh\,\eta}{\sinh\left(z-z_i\right)}\prod_{k\neq i}^{m}\frac{\sinh(z_i-z_k+\eta)}{\sinh(z_i-z_k)}\label{eq:BEoffshellD}.
\end{align}

Substituting the equations \eqref{eq:BEoffshellA} and \eqref{eq:BEoffshellD} in the equation \eqref{eq:MD=-MA} we obtain the famous \textbf{Bethe equations}

\begin{equation}
\left(\frac{\sinh\left(z_i+\eta\right)}{\sinh z_i}\right)^N=\prod_{k\neq i}^{m}\frac{\sinh\left(z_i-z_k+\eta\right)}{\sinh\left(z_i-z_k-\eta\right)} \quad \text{for} \quad i=1,...,m,
\label{eq:BE}
\end{equation}
where $ \{z_k\} $ are called \textbf{Bethe roots}. 

We can rewrite \eqref{eq:eigenvalue} and\eqref{eq:BE} in a more symmetric form by redefining $ z_j $ as $ z_j= \bar{z}_j-\frac{\eta}{2} $ like

\begin{equation}
\Lambda(\bar{z},\{\bar{z}_1,...,\bar{z}_m\})=\prod_{i=1}^{m}\left(\sinh^N\left(\bar{z}+\frac{\eta}{2}\right)\frac{\sinh(\bar{z}-\bar{z}_i-\eta)}{\sinh(\bar{z}-\bar{z}_i)}+\sinh^N\left(\bar{\bar{z}}-\frac{\eta}{2}\right)\frac{\sinh(\bar{z}-\bar{z}_i+\eta)}{\sinh(\bar{z}-\bar{z}_i)}\right)
\label{eq:eigenvaluesymmetric}
\end{equation}
and
\begin{equation}
\left(\frac{\sinh\left(\bar{z}_i+\frac{\eta}{2}\right)}{\sinh \left(\bar{z}_i-\frac{\eta}{2}\right)}\right)^N=\prod_{k\neq i}^{m}\frac{\sinh\left(\bar{z}_i-\bar{z}_k+\eta\right)}{\sinh\left(\bar{z}_i-\bar{z}_k-\eta\right)} \quad \text{for} \quad i=1,...,m.
\label{eq:BEsymmetric.}
\end{equation}
\begin{tcolorbox}
	
	\paragraph{Exercise 24:} Now that we have all the eigenvalues \eqref{eq:eigenvalue} for the transfer matrix of XXZ periodic spin chain, use them to compute the energy in terms of the Bethe roots.
	
\end{tcolorbox}

\begin{tcolorbox}
	
	\paragraph{Exercise 25:} Using Mathematica, for the XXZ spin chain described above
	
	\textbf{a)} compute the eigenvalues of the transfer matrix for  $ N=1,\,2,\,3 $ by direct diagonalization for some numerical values of $ z $ and $ \eta $;\\
	\textbf{b)} using the same numerical values of $ z $ and $ \eta $ solve the Bethe equations \eqref{eq:BE} and check that substituting the Bethe roots in the eigenvalue equation \eqref{eq:eigenvalue} you obtain the same eigenvalues as you obtained by direct diagonalization.
	
\end{tcolorbox}

\begin{tcolorbox}
	
	\paragraph{Exercise 26:} Notice that when we defined $ \cL_{a,j}(z) $ we could have written $ \cL_{a,j}(z-\theta_j) $ instead, where $ \theta_j $ is a parameter usually called inhomogeneity. This would still satisfy the RLL equation \eqref{eq:RLL} and the transfer matrix constructed using this satisfies $ \left[t(z_1,\{\theta_j\}),t(z_2,\{\theta_j\})\right] =0$. So, for XXZ
	
	\textbf{a)} Construct the monodromy matrix
	\begin{equation}
	\mathcal{T}_\alpha(z,\{\theta_j\})=\mathcal{L}_{\alpha,N}(z-\theta_N)\mathcal{L}_{\alpha,N-1}(z-\theta_{N-1})\,...\,\mathcal{L}_{\alpha,1}(z-\theta_1)
	\end{equation}
	and check that the RTT relation holds if we put the same inhomogeneities in both $ \mathcal{T}_{\alpha} $ and $ \mathcal{T}_{\beta} $.
	
	\textbf{b)} Construct the transfer matrix 
	\begin{equation}
	t(z,\{\theta_j\})=\tr_\alpha \mathcal{T}_\alpha(z,\{\theta_j\})
	\end{equation}
	and check that 
	\begin{equation}
	\left[t(z_1,\{\theta_j\}),t(z_2,\{\theta_j\})\right] =0.
	\end{equation}
	\textbf{c)} Now let us see what changes the inhomogeneities cause in the Bethe ansatz. Start by identifying
	
	\begin{equation}
	\mathcal{T}_\alpha(z,\{\theta_j\})=\begin{pmatrix}
	\mathcal{A}(z;\left\{\theta_j\right\}) & \mathcal{B}(z;\left\{\theta_j\right\})\\
	\mathcal{C}(z;\left\{\theta_j\right\}) & \mathcal{D}(z;\left\{\theta_j\right\})
	\end{pmatrix}
	\end{equation}
	and compute the action of $ \mathcal{A}(z;\left\{\theta_j\right\}) $, $ \mathcal{B}(z;\left\{\theta_j\right\}) $, $ \mathcal{C}(z;\left\{\theta_j\right\}) $ and $ \mathcal{D}(z;\left\{\theta_j\right\}) $ on $ |0\rangle $.
	Computing $ N=1,2,3 $ is enough to guess the general formula.
	
	\textbf{d)} Argue that the equation for the eigenvalues \eqref{eq:eigenvalue} and the Bethe equations are modified only by 
	\begin{align}
	& \sinh^N(z+\eta)\mapsto (-1)^N\prod_{j=1}^{N}\sinh(z-\theta_j+\eta),\\
	& \sinh^N(z)\mapsto (-1)^N\prod_{j=1}^{N}\sinh(z-\theta_j).
	\end{align}
	
	\textbf{e)} Finally, write the equations in a symmetric way, like we did for the homogeneous case.
	
\end{tcolorbox}

Although the construction presented in section \ref{subsec:transfer} is general and works for any integrable model, the algebraic Bethe ansatz as presented here does not apply directly to Hilbert spaces of higher dimensions. For such cases, a generalization called Nested algebraic Bethe ansatz (NABA) needs to be applied (for a review of NABA see \cite{Levkovich-Maslyuk:2016kfv} or \cite{Slavnov:2019hdn}). There are other possibilities, for example, if one is interested in the eigenvalues but not in the eigenvectors, the analytical Bethe ansatz is a great option and it can be applied for Hilbert space of any dimension. I am not aware of a review for analytical Bethe ansatz, but for the trigonometric R-matrices in \cite{Jimbo:1985ua}, for example, for periodic spin chains, it was constructed in \cite{Reshetikhin:1987}, while for open spin chains with the most general diagonal boundary matrices it was constructed in \cite{Nepomechie:2018nvl}. 
 
\subsection{R-matrix as an S-matrix}\label{subsec:Smatrix}

Let us now, as promised in the previous sections, quickly provide an interpretation for the R-matrix and the spectral parameters $ z_i $. In short, the R-matrix can be interpreted as a two-body S-matrix in (1+1) quantum field theory particle scattering, while the spectral parameters can be interpreted as momenta (or rapidities) of the particles.

When we consider a (1+1) quantum field theory and we require it to be integrable, we have an infinite set of conserved charges. The presence of such charges implies that the S-matrix has to satisfy the so-called  factorized scattering condition, severely restricting which interactions can happen in these theories \cite{ZAMOLODCHIKOV1979253}.

The consequences of integrability are the following:

\begin{itemize}
	\item There is no particle production and no particle annihilation;
	\item The set of momenta is conserved during the scattering, i.e., $ \left\{p_i\right\} =\left\{p_f\right\}$;
	\item The scattering of $ n $ particles is factorized into a sequence of $ 2\rightarrow 2 $ particles scattering events. This allows us to say that an $ n $-particles scattering can be completely described by the S-matrix $ S_{i,j}(p_i,p_j) $, where $ p_i $ and $ p_j $ are the momenta of the particles $ i $ and $ j $ respectively. 
\end{itemize}

Notice however that this introduces some ambiguity. Consider the scattering of three particles labeled 1, 2 and 3. We can either scatter particles (1,2) first, then (1,3), then (2,3). Or we can scatter (2,3), then (1,3), then (1,2). In order to be consistent we have to require that these two sequences of scattering events give the same result. This means that the two-body S-matrix satisfies the Yang-Baxter equation

\begin{equation}
S_{1,2}(p_1,p_2)S_{1,3}(p_1,p_3)S_{2,3}(p_2,p_3)=S_{2,3}(p_2,p_3)S_{1,3}(p_1,p_3)S_{1,2}(p_1,p_2)
\end{equation} 
which can be graphically represented as

\begin{figure}[H]
	\hspace{4cm}	\includegraphics[width=8cm]{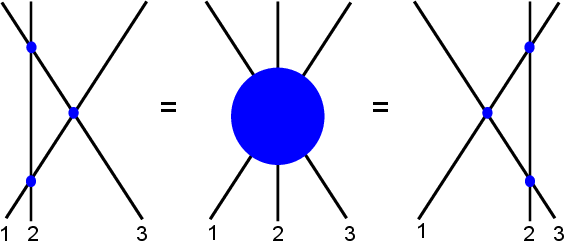}
	\caption{Graphical representation of the Yang-Baxter for the S-matrix. We can interpret each time the lines cross as scattering event described by a two-body S-matrix.}
\end{figure}

\section{Some applications}\label{sec:new}

Integrability is a fascinating topic and in this section we list a selection of more specialized topics that are naturally linked to these notes in case somebody is interested to read more about. This list is incomplete and when available, we refer to reviews and introductions on the subject at hand, so many important references may not be mentioned. The first areas we would like to mention are on the topics discussed in this school:

\begin{itemize}
	\item In order to learn about \textit{Integrable deformations of sigma models} see Ben Hoare's lecture notes  \cite{Hoare:2021dix} and references there in;
	\item In order to learn about \textit{4-dimensional Chern-Simons and integrable field theories} see Sylvain Lacroix's lecture notes \cite{Lacroix:2021iit} and references there in. 
\end{itemize}

Some other interesting areas are:

\begin{itemize}
	\item The very important question of the thermodynamics of integrable models is addressed by the so-called Thermodynamic Bethe ansatz (TBA). In order to learn more about it see \cite{Bajnok:2010ke} and \cite{vanTongeren:2016hhc}. This originated the T-Q construction in AdS, and to learn more about this one can read the reviews in the next item.
	\item Quantum Spectral Curve (QSC)\cite{Gromov:2013pga} is a modern technique in integrability that allows one to compute the exact spectrum of planar $ \mathcal{N}=4 $ Super Yang-Mills theory. In order to learn, see one of the following reviews on this subject \cite{Gromov:2017blm,Levkovich-Maslyuk:2019awk}. The QSC is an alternative to the TBA in the case of planar $ \mathcal{N}=4 $ SYM theory;
	\item A lot of interesting research continues to be made in many aspects of integrability in AdS/CFT. Some interesting places to start in the subject are \cite{Beisert:2010jr,Arutyunov:2009ga};
	\item Separation of variables (SOV) is also a very interesting topic to take a look (see \cite{Sklyanin:1991ss,Sklyanin:1995bm,Kazama:2013rya});
	\item A lot of work has been made in Yangians (see the review \cite{Loebbert:2016cdm}) and Quantum groups (see for example \cite{Pasquier:1989kd,Kulish:1991np}); 
	\item One-point functions in AdS/CFT have also been a very active topic in the recent years as well (for some reviews see \cite{deLeeuw:2017cop,deLeeuw:2019usb} and \cite{Linardopoulos:2020jck});
	\item In these notes we discussed only periodic spin chains. A lot of work has been made in open spin chains and Bethe ansatz as well (two good places to start are \cite{Sklyanin:1988yz,Mezincescu:1990uf});
	\item A new method to find new solutions of qYBE has been proposed in \cite{DeLeeuw:2019gxe,deLeeuw:2020ahe}. We added a review of this method in appendix \ref{app:newmodels};
	\item Lindblad systems have also been recently studied in the context of integrability\cite{Medvedyeva_2016,10.21468/SciPostPhys.8.3.044,deLeeuw:2021cuk};

	\item Integrable defects and Bäcklund transformations have been also extensively studied (see for example \cite{Bowcock:2004my} and references therein);
	\item Very recently the study of set-theoretic YBE got some attention as well \cite{Doikou:2020bhy}.
\end{itemize}

In addition to the ones referred above we would like also to suggest a few reviews/introductions/books in general topics in integrability that we believe could be very useful: \cite{Faddeev:1996iy,Nepomechie:1998jf,Doikou:2009xq,Torrielli:2016ufi,Franchini:2016cxs,Levkovich-Maslyuk:2016kfv,Chari:1994pz,korepin_bogoliubov_izergin_1993,FADDEEV19781,Dorey:1996gd,essler2005one}.

\section*{Acknowledgments}

I would like to thank Saskia Demulder and Fedor Levkovich-Maslyuk very much for carefully reading the manuscript and giving many valuable suggestions. I am also very grateful to Riccardo Borsato, Saskia Demulder, Sibylle Driezen, Fedor Levkovich-Maslyuk and Emanuel Malek for the all the support before and during the School. I would also like to thank Rafael Nepomechie and Marius de Leeuw for advises, and Ben Hoare, Sylvain Lacroix, Marius de Leeuw, Paul Ryan, Anton Pribytok and Chiara Paletta for valuable discussions.   These notes were written for the lectures delivered at the school ``Integrability, Dualities and Deformations", that ran from 23 to 27 August 2021 in Santiago de Compostela and virtually \url{https://indico.cern.ch/e/IDD2021}. I am partially supported by Grant No. 18/EPSRC/3590. 
  
\appendix

\section{Proof of equation \eqref{eq:PoissonL1L2}}\label{app:proof}

In this section we prove (based on the discussion in \cite{babelon_bernard_talon_2003}) the equation \eqref{eq:PoissonL1L2}  given by 

\begin{equation}
\left\{L_1,L_2\right\}=\left[r_{12},L_1\right]-\left[r_{21},L_2\right].
\end{equation}

The first step is to substitute $ L $ by $ U\Lambda U^{-1} $ as we see below
\begin{align}
\left\{L_1,L_2\right\}& =\left\{U_1\Lambda_1 U_1^{-1}, U_2\Lambda_2 U_2^{-1}\right\}\label{eq:PoissonL1L2app1}\\
& =\left\{U_1,U_2\right\}\Lambda_2U_2^{-1}\Lambda_1U_1^{-1}+U_2\Lambda_2\left\{U_1,U_2^{-1}\right\}\Lambda_1U_1^{-1}+\nonumber\\
&\quad+U_1\Lambda_1\left\{U_1^{-1},U_2\right\}\Lambda_2U_2^{-1}+U_1U_2\Lambda_1\Lambda_2 \left\{U_1^{-1},U_2^{-1}\right\}+\nonumber\\
&\quad+U_2\left\{U_1,\Lambda_2 \right\}U_2^{-1}\Lambda_1U_1^{-1}+U_1\left\{\Lambda_1,U_2 \right\}\Lambda_2U_2^{-1}U_1^{-1}+\nonumber\\
&\quad+U_1U_2\Lambda_2\left\{\Lambda_1,U_2^{-1}\right\}U_1^{-1}+U_1U_2\Lambda_1\left\{U_1^{-1},\Lambda_2\right\}U_2^{-1}+\nonumber\\
&\quad +U_1U_2\left\{\Lambda_1,\Lambda_2 \right\}U_1^{-1}U_2^{-1}.\label{eq:PoissonL1L2app2}
\end{align}
To go from \eqref{eq:PoissonL1L2app1} to \eqref{eq:PoissonL1L2app2} we use a few times properties (4) and (5) from section \ref{subsec:Poissonbracket}, and also use that operators acting on different sites commute.

The last term is zero because $\left\{\Lambda_1,\Lambda_2 \right\}=0 $.

The four first terms depend on $ \left\{U_1,U_2\right\} $, so let us deal with them first. The first term is given by

\begin{align}
\left\{U_1,U_2\right\}\Lambda_2U_2^{-1}\Lambda_1U_1^{-1}&=\left\{U_1,U_2\right\}U_1^{-1}U_1U_2^{-1}U_2\Lambda_2U_2^{-1}\Lambda_1U_1^{-1}\nonumber\\
&=\left(\left\{U_1,U_2\right\}U_1^{-1}U_2^{-1}\right)\left(U_2\Lambda_2U_2^{-1}\right)\left(U_1\Lambda_1U_1^{-1}\right)\nonumber\\
&=k_{12}\,L_1\,L_2
\end{align}
where from the first to the second line we use that operators acting on different sites commute, and in the second line we define $ k_{12}=\left\{U_1,U_2\right\}U_1^{-1}U_2^{-1} $.

The second term is then written as

\begin{align}
U_2\Lambda_2\left\{U_1,U_2^{-1}\right\}\Lambda_1U_1^{-1}&=-U_2\Lambda_2U_2^{-1}\left\{U_1,U_2\right\}U_2^{-1}\Lambda_1U_1^{-1}\nonumber\\
&=-L_2\left(\left\{U_1,U_2\right\}U_1^{-1}U_2^{-1}\right)(U_1\Lambda_1U_1^{-1})\nonumber\\
&=-L_2k_{12}L_1
\end{align}
where we use the equation \eqref{eq:Poissoninverse}, the commutativity of operators acting on different spaces and the definition of $ k_{12} $ again.

Similarly, the third and fourth terms give

\begin{equation}
U_1\Lambda_1\left\{U_1^{-1},U_2\right\}\Lambda_2U_2^{-1}=-L_1k_{12}L_2,
\end{equation}
and

\begin{equation}
U_1U_2\Lambda_1\Lambda_2 \left\{U_1^{-1},U_2^{-1}\right\}=L_1L_2k_{12},
\end{equation}
respectively.

So, the four first terms give 

\begin{equation}
\left[\left[k_{12},L_2\right],L_1\right],
\end{equation}
which can be rewritten as
\begin{equation}
\left[\left[k_{12},L_2\right],L_1\right]=\frac{1}{2}\left[\left[k_{12},L_2\right],L_1\right]-\frac{1}{2}\left[\left[k_{21},L_1\right],L_2\right].
\end{equation}

Now, to the remaining four terms. The fifth term becomes

\begin{align}
U_2\left\{U_1,\Lambda_2 \right\}U_2^{-1}\Lambda_1U_1^{-1}&=\left(U_2\left\{U_1,\Lambda_2 \right\}U_2^{-1}U_1^{-1}\right)\left(U_1\Lambda_1U_1^{-1}\right)\nonumber\\
&=q_{12}L_1
\end{align}
where $ q_{12}=U_2\left\{U_1,\Lambda_2 \right\}U_2^{-1}U_1^{-1} $.

Similarly, the remaining three terms are given by
\begin{align}
& U_1\left\{\Lambda_1,U_2 \right\}\Lambda_2U_2^{-1}U_1^{-1}=-q_{21}L_2\\
&U_1U_2\Lambda_2\left\{\Lambda_1,U_2^{-1}\right\}U_1^{-1}=L_2q_{21}\\
&U_1U_2\Lambda_1\left\{U_1^{-1},\Lambda_2\right\}U_2^{-1}=-L_1q_{12}
\end{align}
So, these four terms give

\begin{equation}
\left[q_{12},L_1\right]-\left[q_{21},L_2\right].
\end{equation}

Finally, putting all the terms together 

\begin{equation}
\left\{L_1,L_2\right\}=\left[q_{12}+\frac{1}{2}\left[k_{12},L_2\right],L_1\right]-\left[q_{21}+\frac{1}{2}\left[k_{21},L_1\right],L_2\right],
\end{equation}
by defining
\begin{equation}
r_{12}=	q_{12}+\frac{1}{2}\left[k_{12},L_2\right]
\end{equation}
we can write 
\begin{equation}
\left\{L_1,L_2\right\}=\left[r_{12},L_1\right]-\left[r_{21},L_2\right].
\end{equation}

\textbf{}

\section{Constructing the Lax pair for integrable hierarchies}\label{app:Lax}

There are many examples of integrable hierarchies in the literature, to cite a few: the KdV, AKNS, mKdV, Drinfeld-Sokolov and KP hierarchies. Not all of them, however, fit in the construction we present in this appendix. Our main goal here is just to give a glimpse of such constructions and show explicitly an example. 

The construction of integrable hierarchies depends on two choices\footnote{In the way I am presenting it here. There are different ways to do it.}: the algebra and the gradation. Usually for a given hierarchy, the $ \cL_{x} $ is the same for the whole hierarchy, while $ \cL_{t_M} $ is different for each equation ($ M \in \mathbb{Z}$ is a label that specifies in which equation of the hierarchy we are). Both are constructed in order to satisfy the zero curvature equation.

\begin{equation}
\left[\partial_x+\cL_{x},\partial_{t_M}+\cL_{t_M}\right]=0
\label{eq:zc}
\end{equation}

\subsection{General concepts}\label{subapp:general}
\subsubsection{The algebra}\label{subsubapp:algebra}
In this appendix we will focus only on the Kac-moody algebra $ \widehat{sl}(2) $ which is composed by the following commutation relations

\begin{align}
& \left[h^{(m)},h^{(n)}\right]=2m\delta_{m+n,0}\hat{c},\\
& \left[h^{(m)},e_{\pm\alpha}^{(n)}\right]=\pm 2 e_{\pm \alpha}^{(m+n)},\\
& \left[e_{\alpha}^{(m)},e_{\beta}^{(n)}\right]=\begin{cases}
 h^{(m+n)}+m\delta_{m+n,0}\hat{c} & \text{for $\alpha+\beta=0$ }\\
0 & \text{otherwise}
\end{cases}, \\
& \left[\hat{c},e_{\pm \alpha}^{(m)}\right]=\left[\hat{c},h^{(m)}\right]=\left[\hat{c},\hat{d}\right]=0,\\
& \left[\hat{d},e_{\pm \alpha}^{(m)}\right]=m e_{\pm \alpha}^{(m)},\\
& \left[\hat{d},h^{(m)}\right]=m h^{(m)},
\end{align}
where $ m,n\in \mathbb{Z} $.

In order to construct the hierarchy, the central terms $ \hat{c} $ are not relevant (although they are very important if we want to solve the equations in the hierarchy), so we will consider $ \hat{c}=0 $. 
Also, we will consider 

\begin{align}
& h^{(n)}=z^n h^{(0)},\\
& e_{\pm\alpha}^{(n)}=z^n e_{\pm\alpha}^{(0)}
\end{align}
and 

\begin{equation}
\hat{d}=z\frac{d}{dz}.
\end{equation}

\subsubsection{The gradation}\label{subsubapp:gradation}

The \textbf{gradation operator $ \hat{G} $} is an operator which decomposes the algebra in graded subspaces

\begin{equation}
\hat{g}=\oplus \hat{g}^{(n)}
\end{equation}
where

\begin{equation}
\left[\hat{G}^{(m)},\hat{G}^{(n)}\right]\subset \hat{G}^{(m+n)}.
\end{equation}

The algebra $ \widehat{sl}(2) $ has two main gradations

\begin{itemize}
	\item the \textbf{homogeneous gradation} described by the operator
	\begin{equation}
	G=z \frac{d}{dz}=\hat{d},
	\end{equation}
	\item and the \textbf{principal gradation} described by the operator
	\begin{equation}
	G=2z\frac{d}{dz}+\frac{1}{2}h^{(0)}=2\hat{d}+\frac{1}{2}h^{(0)}.
	\end{equation}
\end{itemize}

\subsubsection{The strategy}\label{subsubapp:strategy}

The strategy consists in using the concept of gradation to decompose the Lax connection in a way that we can find the hierarchy equations. For that, we define

\begin{equation}
\cL_x(x,t_M)=\cL_0(x,t_M)+E^{(1)}
\end{equation}
where $ \cL_0 $ is the object that contains the fields of the model and it has degree zero, while
$ E^{(1)} $ is a constant operator and has degree one. The explicit form of them will depend if we are in the homogeneous or in the principal gradation as we will see below. 

Now about $ \cL_{t_M} $. It can be decomposed in two different ways depending if $ M $ is positive or negative.

For positive $ M $ 

\begin{equation}
\cL_{t_M}=D^{(M)}+D^{(M-1)}+...+D^{(1)}+D^{(0)}, 
\label{positivedecomposition}
\end{equation}
where $ D^{(i)} $ is the most general ansatz (in terms of the generators) with degree $ i $. The index $ M $ in $ t_M $ is just to indicate which equation in the hierarchy we are referring to. With this definition the zero curvature equation can be written like

\begin{equation}
\left[\partial_x+\cL_0+E^{(1)},\partial_{t_M}+D^{(M)}+D^{(M-1)}+...+D^{(1)}+D^{(0)}\right]=0.
\end{equation}

We can then decompose this equation in degrees:

\begin{align}
&(M+1):&&   \left[E^{(1)},D^{(M)}\right] =0\label{higherdegree},\\
&(M): && \partial_xD^{(M)}+\left[\cL_0,D^{(M)}\right]+\left[E^{(1)},D^{(M-1)}\right]=0,\\
&\vdots && \hspace{2cm} \vdots\nonumber\\
&(1):  &&\partial_xD^{(1)}+\left[\cL_0,D^{(1)}\right]+\left[E^{(1)},D^{(0)}\right]=0,\\
&(0):  &&\partial_xD^{(0)}+\left[\cL_0,D^{(0)}\right]-\partial_{t_M}D^{(0)}=0\label{smallestdegree}.
\end{align}

Notice that the only equation depending on $ \partial_{t_M} $ is the one with degree zero. So the strategy is to start by solving the equation with higher degree to find $ D^{(M)} $, then substitute it in the next one to find $ D^{(M-1)} $, and we keep going until we solve the equation of degree zero and find the equations of motion. Notice that if we construct all the $ D^{(i)} $ we have the Lax connection as a Laurent expansion in the spectral parameter $ z $. We will soon see an example and things will hopefully become clearer. 

If we now write the decomposition for $ M<0 $ we have 

\begin{equation}
\cL_{t_M}=D^{(M)}+D^{(M+1)}+...+D^{(-2)}+D^{(-1)}, 
\label{negativedecomposition}
\end{equation}
where again $ D^{(i)} $ is the most general ansatz (in terms of the generators) with degree $ i $. With this definition the zero curvature equation can be written like

\begin{equation}
\left[\partial_x+\cL_0+E^{(1)},\partial_{t_M}+D^{(M)}+D^{(M+1)}+...+D^{(-2)}+D^{(-1)}\right]=0
\end{equation}
whose decomposition in degrees is 

\begin{align}
& (M): && \partial_xD^{(M)}+\left[\cL_0,D^{(M)}\right]=0,\\
& (M+1): && \partial_xD^{(M+1)}+\left[\cL_0,D^{(M+1)}\right]+\left[E^{(1)},D^{(M)}\right]=0,\\
&\vdots && \hspace{2cm} \vdots\nonumber\\
& (-1): &&\partial_xD^{(-1)}+\left[\cL_0,D^{(-1)}\right]+\left[E^{(1)},D^{(-2)}\right]=0,\\
& (0): && \left[E^{(1)},D^{(-1)}\right]-\partial_{t_M}\cL_0=0.
\end{align}
Again, the strategy is to solve first the equation most further from degree 0 and keep solving one by one until reaching the equation for degree zero which will give us the equation of motion. 

This all sounds a bit abstract, so let us study two examples. 

\subsection{Examples}\label{subapp:examples}

We will focus on two examples with the principal gradation (see also  \cite{Gomes_2009}). So, if we take the principal gradation operator and act with it on the generators of $ \widehat{sl}(2) $ we obtain the following
\begin{align}
&\left[G,h^{(m)}\right]=2\,m\,h^{(m)},\\
&\left[G,e_{\pm\alpha}^{(m)}\right]=(2m\pm 1)e_{\pm \alpha}^{(m)}.
\end{align}
\noindent
This means that for this choice of gradation, the set of operators with even degree is given by
\begin{equation}
g_{2m}=\left\{h^{(m)}\right\}
\label{evendegrees}
\end{equation}
and the set of operators with odd degree is given by

\begin{equation}
g_{2m\pm 1}=\left\{e_{\alpha}^{(m)},e_{-\alpha}^{(m+1)}\right\}.
\label{odddegrees}
\end{equation}

With this it is easy to see that the most general operator of degree 0, is proportional to $ h^{(0)} $, so we can define 

\begin{equation}
\cL_0=v(x,t)h^{(0)}
\label{L0}
\end{equation}
and for degree one is composed by a linear combination of $ e_{\alpha}^{(0)} $ and $ e_{-\alpha}^{(1)} $

\begin{equation}
E^{(1)}=e_{\alpha}^{(0)}+e_{-\alpha}^{(1)}.
\label{E1}
\end{equation}
Once we fixed these objects that compose $ \cL_x $ they will be the same for the whole hierarchy. This hierarchy is called the \textbf{mKdV hierarchy}. Each $ M $ defines a new equation inside this hierarchy. Let us then see how to explicitly construct $ \cL_{t_M} $ for two examples.

\subsubsection{Sinh-Gordon model, $ M=-1 $}

If we consider $ M=-1 $ in the negative decomposition we have two equations

\begin{align}
& (-1): &&\partial_xD^{(-1)}+\left[\cL_0,D^{(-1)}\right]+\left[E^{(1)},D^{(-2)}\right]=0,\label{degree-1}\\
& (0): && \left[E^{(1)},D^{(-1)}\right]-\partial_{t_{-1}}\cL_0=0\label{degree0}.
\end{align}

The first step now is to write an ansatz for $ D^{(-1)} $. A good starting ansatz is to take the most general linear combination of generators with degree -1. By looking into equation \eqref{odddegrees} we see that such generators are $ e_{\alpha}^{(-1)} $ and $ e_{-\alpha}^{(0)} $, and we can write

\begin{equation}
D^{(-1)}=a\,e_{\alpha}^{(-1)}+b\,e_{-\alpha}^{(0)}.
\end{equation}
Substituting this $ D^{(-1)} $, $ E^{(1)} $ (as in equation \eqref{E1}), and $ \cL_0 $ (as in equation \eqref{L0}) in equation \eqref{degree-1} and performing the commutation relations we find

\begin{equation}
\left(\partial_{x}a+2av\right)e_{\alpha}^{(-1)}+\left(\partial_xb-2bv\right)e_{-\alpha}^{(0)}=0.
\end{equation}
Now, each of the terms have to be zero so, solving them we obtain

\begin{equation}
a=e^{-2\int v\, dx} \quad \text{and} \quad b=e^{2\int v dx}.
\end{equation}

So, 

\begin{equation}
D^{(-1)}=e^{-2\int v\, dx}\,e_{\alpha}^{(-1)}+e^{2\int v dx}\,e_{-\alpha}^{(0)}.
\end{equation}
Substituting this in equation \eqref{degree0} we obtain that

\begin{equation}
\partial_{t_{-1}} v=2\sinh\left( 2\int v\, dx\right)
\end{equation}
which becomes the \textbf{Sinh-Gordon equation} if we assume $ v=\partial_x \phi $:
\begin{equation}
\partial_{t_{-1}} \partial_x \phi=2\sinh\left( 2\phi\right).
\end{equation}
Notice that these are light-cone variables, so $ x $ is actually $ x^+ $ and $ t_{-1}= x^- $. 

\subsubsection{mKdV equation, $ M=3 $}\label{subsecapp:mKdV}

Consider the positive decomposition \eqref{positivedecomposition} with $ M=3 $

\begin{align}
&(4):&&   \left[E^{(1)},D^{(3)}\right] =0,\label{mkdVdegree4}\\
&(3): && \partial_xD^{(3)}+\left[\cL_0,D^{(3)}\right]+\left[E^{(1)},D^{(2)}\right]=0,\label{mkdVdegree3}\\
&(2):  &&\partial_xD^{(2)}+\left[\cL_0,D^{(2)}\right]+\left[E^{(1)},D^{(1)}\right]=0,\label{mkdVdegree2}\\
&(1):  &&\partial_xD^{(1)}+\left[\cL_0,D^{(1)}\right]+\left[E^{(1)},D^{(0)}\right]=0,\label{mkdVdegree1}\\
&(0):  &&\partial_xD^{(0)}+\left[\cL_0,D^{(0)}\right]-\partial_{t_M}D^{(0)}=0\label{mkdVdegree0}.
\end{align}

The first step is to use equations \eqref{evendegrees} and \eqref{odddegrees} and write the ansatz for $ D^{(i)} $'s

\begin{align}
& D^{(3)}=a_3\,e_{\alpha}^{(1)}+b_3\,e_{-\alpha}^{(2)},\label{D3}\\
& D^{(2)}=c_2\,h^{(1)},\label{D2}\\
& D^{(1)}=a_1\,e_{\alpha}^{(0)}+b_1\,e_{-\alpha}^{(1)},\label{D1}\\
& D^{(0)}=c_0\,h^{(0)},\label{D0}
\end{align}
where the coefficients $ a_i,\,b_i,\, c_i $ until this moment are arbitrary functions of $ x $ and $ t_3 $. We will work now on constructing their explicit form.

Remember now, the strategy is to start with the equation of higher degree and solving one by one until the one with degree 0. 

Substituting $ D^{(3)} $ in equation \eqref{mkdVdegree4} we obtain that $ D^{(3)} $ is proportional to $ E^{(1)} $, i.e., 

\begin{equation}
D^{(3)}=a_3\left(e_{\alpha}^{(1)}+e_{-\alpha}^{(2)}\right).
\end{equation}
We can then substitute this, together with $ E^{(1)} $, $ \cL_0$ and the ansatz for $ D^{(2)} $ in  equation \eqref{mkdVdegree3}. By performing all the commutation relations one finds

\begin{equation}
a_3=\text{constant}\equiv \alpha, \quad \text{and}\quad c_2=\alpha  v .
\end{equation}

Repeating this procedure until you reach the equation with degree zero we obtain the complete picture

\begin{align}
& D^{(3)}=\alpha\left(\,e_{\alpha}^{(1)}+\,e_{-\alpha}^{(2)}\right),\label{D3ready}\\
& D^{(2)}=\alpha\,v\,\,h^{(1)},\label{D2ready}\\
& D^{(1)}=\frac{1}{2}\left(\partial_xv-v^2\right)\,e_{\alpha}^{(0)}-\frac{1}{2}\left(\partial_xv+v^2\right)\,e_{-\alpha}^{(1)},\label{D1ready}\\
& D^{(0)}=\left(\frac{1}{4}\partial^2_xv-\frac{v^3}{2}\right)\,h^{(0)}\label{D0ready}
\end{align}
which gives the so called \textbf{ mKdV equation}

\begin{equation}
4\,\partial_{t_3}v=\partial^3_xv-6\,v\,\partial_xv.
\end{equation} 

For positive hierarchy we started with $ M=3 $. One could ask why not start with a simpler example like $ M=1 $ or $ M=2 $. The answer is that $ M=1 $ is trivial, while $ M=2 $ is not allowed. It so happens that actually for $ M>0 $ only odd values of $ M $ are allowed. We will leave the reason why this happens as an exercise.

\begin{tcolorbox}
	
	\paragraph{Exercise: 27} Understand why for $ M>0 $ only odd $ M $ makes sense for the mKdV hierarchy.
	In order to do that, try to follow the above procedure for $ M=2 $, for example. Explain why the same problem does not happen for the negative part of the hierarchy.
	
\end{tcolorbox}

\subsection{The AKNS hierarchy}\label{subapp:AKNS}

What happens if we consider the homogeneous gradation instead of the principal gradation?

Since the homogeneous gradation operator is just $ \hat{d} $, the generators with degree $ m $ are given by 

\begin{equation}
g_{m}=\left\{ e_{\alpha}^{(m)},\,e_{-\alpha}^{(m)},\,h^{(m)}  \right\}.
\label{homogeneousgm}
\end{equation}
Since the degree is always equal to the upper index of all the generators, the calculations are much simpler than with the principal gradation. 

In order to construct the AKNS hierarchy we assume

\begin{align}
&\cL_0=q_1\,e_{\alpha}^{(0)}+q_2\,e_{-\alpha}^{(0)},\\
&E^{(1)}=h^{(1)}
\end{align}
where $ q_1\equiv q_1(x,t_M) $ and $ q_2\equiv q_2(x,t_M) $ are the fields of the model.

\begin{tcolorbox}
	\paragraph{Exercise 28:} In this exercise you will construct the first nontrivial equations in the AKNS hierarchy, which are obtained for $ M=2 $:
	
	\textbf{a)} Consider $ M=2 $,  write the ansatz for $ D^{(2)} $, $ D^{(1)} $ and $ D^{(0)} $ using \eqref{homogeneousgm};
	
	\textbf{b)} Substitute these ansatz in the positive decomposition \eqref{higherdegree}-\eqref{smallestdegree} for $ M=2 $ and solve to find the two equations that describe this system for $ M=2 $;
	
	\textbf{c)} Check that by assuming $ q_1=\psi $ and $ q_2 =\psi^*$ you obtain the Nonlinear Schrödinger equation (NLS) discussed in exercise 13.
\end{tcolorbox}

As already mentioned at the beginning of this appendix, an important comment is that not all hierarchies fit in the framework described above. In addition to the references already cited, see for example, \cite{Lacroix:2018njs,Fioravanti:1998ha,Gomes:2006en} in order to learn more about integrable hierarchies.

Also, there are many subtleties related to the negative part of the hierarchy that are not being addressed here. The negative flows behave in a rather different way than the positive flows, being the main difference related to non-locality. In order to understand  more about this see \cite{Fioravanti:1998uv,Fioravanti:1999th,Fioravanti:2000fu}.

\section{New models}\label{app:newmodels}

There are many ways to solve the quantum Yang-Baxter equation \eqref{eq:qYBEnondif} and find new integrable models \cite{Kulish:1981gi,Kulish1982,Reshetikhin_1983,Jimbo:1985ua,Bazhanov:1986mu,jones1990baxterization,Idzumi:1994kx,Vieira:2017vnw}. Let us shortly explain a method \cite{deLeeuw:2019zsi,deLeeuw:2020ahe} we have been working on to find new regular solutions of the qYBE.

First of all, let us say that in addition to the method presented in section  \ref{subsec:Qtransfer}, there is another way to construct conserved charges: the Boost operator \cite{Tetelman,HbbBoost,Grabowski:1994rb,Loebbert:2016cdm} formalism. The Boost operator is given by

\begin{equation}
\mathbb{B}\left[\mathbb{Q}_2\right]=\partial_z+\sum_{i=-\infty}^{\infty}i\,H_{i,i+1}(z)
\label{eq:boost}
\end{equation}
and its advantage is that if one starts with an integrable Hamiltonian, it can generate higher conserved charges recursively 

\begin{equation}
\mathbb{Q}_{j+1}\sim \left[\mathbb{B}\left[\mathbb{Q}_2\right],\mathbb{Q}_{j}\right],\quad j>1.
\label{eq:higherboost}
\end{equation}

In principle, equation \eqref{eq:boost} seems to work only for infinite chains, however, equation \eqref{eq:higherboost} is well defined for closed chains as well. This happens because of property \eqref{eq:diffspaces}, i.e., because the extra terms given in  the boost give zero when we perform the commutator in \eqref{eq:higherboost}. With this in mind the method consists in applying the following steps:

\textbf{1)} Start with an ansatz Hamiltonian with four sites, 

\begin{equation}
\mathbb{Q}_2=\mathbb{H}(z)=\sum_{i=1}^{4}H_{i,i+1}(z).
\end{equation}

\textbf{2)} Construct the next conserved charge $ \mathbb{Q}_3(z) $ using the boost operator (also for four sites)

\begin{align}
\mathbb{Q}_3(z)& =\left[\mathbb{B}\left[\mathbb{Q}_2(z)\right],\mathbb{Q}_{2}(z)\right]\nonumber\\
&=\sum_{i=1}^{N}\left[H_{i-1,i}(z),H_{i,i+1}(z)\right]+\frac{d\mathbb{H}(z)}{dz}.
\label{Q3}
\end{align}
Notice that $ \mathbb{Q}_3(z) $ depends only on the ansatz Hamiltonian we started with.

\textbf{3)} Require that 

\begin{equation}
\left[\mathbb{Q}_2(z),\mathbb{Q}_3(z)\right]=0
\label{eq:Q2Q3}
\end{equation}
and solve the differential equations for the matrix elements of $ H_{i,i+1} $. This will give us several (the exact number depends on the ansatz) potentially integrable Hamiltonians. But to be sure they are really integrable we need to ensure that all the charges commute, and not only these two. The way to do that is to solve the Yang-Baxter equation taking the Hamiltonians obtained from equation \eqref{eq:Q2Q3} as boundary conditions.

\textbf{4)} The first step to solve qYBE is to have an ansatz for the R-matrix, i.e., which elements in our R-matrix we expect to be nonzero. In order to decide this, it is useful to notice that the R-matrix can be written as an expansion in the density Hamiltonian

\begin{equation}
R_{1,2}(z_1,z_2) =P_{12}\left(1+(z_1-z_2)H_{12}\left(\frac{z_1+z_2}{2}\right)+O((z_1-z_2)^2)\right).
\end{equation}
Plugging each of the Hamiltonians (solutions of equation \eqref{eq:Q2Q3}) in this expansion we learn which ansatz (i.e. which nonzero elements in R) we should start with to deduce the corresponding R-matrix for each of them.

\textbf{5)} The next step is to solve the qYBE. In order to do that we apply a derivative in the qYBE  with respect to $ z_1 $ and then do $ z_2\rightarrow z_1 $. Using 

\begin{equation}
R(z,z)=P, \quad H(z_1)=P\frac{\partial R(z_1,z_2)}{\partial z_1}\Big|_{z_2\rightarrow z_1}
\label{eq:BC}
\end{equation}
we obtain a Sutherland equation

\begin{equation}
\left[R_{13}(z_1,z_3)R_{23}(z_1,z_3),H_{12}(z_1)\right]=\frac{\partial R_{13}(z_1,z_3)}{\partial z_1}R_{23}(z_1,z_3)-R_{13}(z_1,z_3)\frac{\partial R_{23}(z_1,z_3)}{\partial z_1}.
\label{Sutherland}
\end{equation}
Now, one simply needs to substitute the ansatz for the R-matrix and the Hamiltonian in this equation and solve the differential equations to find the explicit form of $ R(z_1,z_2) $.

\textbf{6)} The last step is to substitute each of these R-matrices in the Yang-Baxter equation to make sure it is satisfied. 

One important detail however, is that step 3 generates many dependent solutions. This is because qYBE remains invariant if the following transformations are performed on the R-matrix:

\begin{itemize}
	\item Normalization:
	\begin{equation}
	R(z_1,z_2)\mapsto g(z_1,z_2)R(z_1,z_2), \quad H(z_1)\mapsto g(z_1,z_1)H(z_1)+\dot{g}(z_1,z_1)I,
	\end{equation}
	\item Reparametrization;
	\begin{equation}
	R(z_1,z_2)\mapsto R(g(z_1),g(z_2)), \quad H(z_1)\mapsto \dot{g}H(g(z_1)),
	\end{equation}
	\item Discrete transformations;
	\begin{align}
	&R(z_1,z_2)\mapsto PR(z_1,z_2)P,\\
	&R(z_1,z_2)\mapsto R^T(z_1,z_2),\\
	&R(z_1,z_2)\mapsto PR^T(z_1,z_2)P,
	\end{align}
	\item Local basis transformations
	\begin{equation}
	R(z_1,z_2)\mapsto \left[V(z_1)\otimes V(z_1)\right]R(z_1,z_2)\left[V(z_1)\otimes V(z_1)\right]^{-1}
	\end{equation}
	which implies
	\begin{equation}
	H(z_1)\mapsto \left(V\otimes V\right)H(z_1)\left(V\otimes V\right)^{-1}-\left[\dot{V}V^{-1}\otimes I-I\otimes \dot{V}V^{-1}\right]
	\end{equation}
	\item Twists:
	\begin{equation}
	R_{12}(z_1,z_2)\mapsto U_2(z_2)R_{12}(z_1,z_2)U_1(z_1)^{-1} \quad \text{if} \quad \left[U(z_1)\otimes U(z_2),R_{12}(z_1,z_2)\right]=0 
	\end{equation}
	which implies
	\begin{equation}
	H_{12}\mapsto U_1H_{12}U_1^{-1}+\dot{U}_1U_1^{-1}\quad \text{if} \quad \left[U_1U_2,H_{12}\right]=\dot{U}_1U_2-U_1\dot{U}_2. 
	\end{equation}
\end{itemize}

So, when we obtain many potentially integrable Hamiltonians in step 3, we compare them to see if any are related through these transformations. We keep only the independent ones. 

\begin{tcolorbox}
	\paragraph{Exercise 29:} Let us apply the method for a simple example. Consider the ansatz Hamiltonian
	\begin{equation}
	H_{1,2}(z)=\begin{pmatrix}
	0 & 0 & 0 & 0\\
	0 & h_1(z) & h_3(z) & 0\\
	0 & h_4(z) & h_2(z) & 0\\
	0 & 0 & 0 & 0
	\end{pmatrix}
	\end{equation}
	\textbf{a)} Construct $ \mathbb{H} $ for 4 sites;\\
	\textbf{b)} Construct $ \mathbb{Q}_3 $ using \eqref{Q3};\\
	\textbf{c)} Solve $ \left[\mathbb{Q}_2,\mathbb{Q}_3\right] =0$ to find $ h_i(z) $;\\
	\textbf{d)} Construct the R-matrix using Sutherland equation \eqref{Sutherland};\\
	\textbf{e)} Check that this R-matrix really satisfies qYBE;
\end{tcolorbox}

This method was successfully applied to construct models that are interesting both in condensed matter and in high energy physics.  For example, the method allowed for a classification of all integrable deformations of AdS$ _3 $ and AdS$ _2 $ R-matrices \cite{deLeeuw:2020ahe,deLeeuw:2020xrw,deLeeuw:2021ufg}. It also provided a systematic way to construct integrable Lindblad superoperators \cite{deLeeuw:2021cuk}. Moreover, it allowed to find many other new models still without a physical interpretation\cite{deLeeuw:2019vdb}.

\bibliographystyle{utphys}
\bibliography{refs}

\end{document}